\documentclass[article,nojss]{jss}

\usepackage{orcidlink,thumbpdf,lmodern}

\usepackage{framed}
\usepackage{caption}
\usepackage{subcaption}
\usepackage{float}
\usepackage{multirow}
\usepackage{microtype}
\usepackage{amsmath, amssymb, bm}
\usepackage{algorithm}
\usepackage{algpseudocode}
\usepackage{caption}
\usepackage{booktabs,array}

\makeatletter
\renewcommand\paragraph{%
  \@startsection{paragraph}{4}{\z@}%
  {0.7ex \@plus 0.2ex \@minus 0.1ex}
  {-1em}
  {\normalfont\normalsize\bfseries}
}
\makeatother

\author{Antoine Poirot-Bourdain~\orcidlink{https://orcid.org/0009-0008-1211-0860
}\\ CNRS, INRIA, INSERM \\ Université Paris-Dauphine PSL \\ Université Paris Cité \\ 
   \And Alexis Van Straaten\\ AP-HP, INRIA, INSERM\\ Université Paris Cité 
   \AND Emmanuel Bacry~\orcidlink{0000-0001-5997-1942}\\CNRS  \\ Université Paris-Dauphine PSL
   \And Anne-Sophie Jannot~\orcidlink{0000-0002-8001-8539}\\ AP-HP, INRIA, INSERM\\ Université Paris Cité 
   \And Jean Feydy~\orcidlink{0000-0001-6049-563X}\\ INRIA, INSERM \\ Université Paris Cité}
\Plainauthor{Antoine Poirot-Bourdain, Alexis Van Straaten, Emmanuel Bacry, Anne-Sophie Jannot, Jean Feydy }
\title{\pkg{survivalGPU}: Scalable Survival Analysis for \proglang{R} and \proglang{Python}}
\Plaintitle{SurvivalGPU: A Scalable Survival Analysis Package}
\Shorttitle{SurvivalGPU: Scalable Survival Analysis}

\Abstract{
This paper introduces \pkg{survivalGPU}, a high-performance \proglang{Python} and \proglang{R} package designed to accelerate survival analysis computations using graphics processing units (GPUs). While the Cox proportional hazards model remains a cornerstone of survival analysis, it can impose substantial computational burdens when applied to large longitudinal data sets. To address this limitation, \pkg{survivalGPU} implements Cox proportional hazards models and weighted cumulative exposure (WCE) models in \proglang{Python}, using \pkg{PyTorch} and \pkg{KeOps} for efficient back-end computation, with interfaces in both \proglang{Python} and \proglang{R}. Numerical validation shows that \pkg{survivalGPU} reproduces the results of reference implementations within the tolerances expected for the selected floating-point precision, with relative errors below $10^{-2}$ in \code{float32} and below $10^{-5}$ in \code{float64}. Benchmark experiments demonstrate substantial speedups. For Cox models, \code{coxphGPU()} was $5.6\times$ faster than \code{coxph()} from the \pkg{survival} package on 1{,}000{,}000 patients, and $14.0\times$ faster based on runtimes extrapolated to 1{,}000 bootstrap replicates on 100{,}000 patients. Gains were larger for WCE models: \code{wceGPU()} was $920.7\times$ faster than \code{WCE()} from the \pkg{WCE} package on 100{,}000 patients, and $1{,}321.4\times$ faster based on runtimes extrapolated to 1{,}000 bootstrap replicates. These results show that \pkg{survivalGPU} makes flexible time-dependent survival modeling and bootstrap-based estimation of confidence intervals feasible for large-scale pharmacoepidemiological applications.
}

\Plainkeywords{R, Python, survival analysis, WCE, Cox model, GPU computing, bootstraps}

\Address{
  Antoine Poirot-Bourdain\\
  CEntre de REcherche en MAthématiques de la DÉcision\\
  CNRS-UMR 7534\\
  Université Paris-Dauphine PSL\\
  75016 Paris, France\\
  \emph{and}\\
  HeKA\\
  Université Paris Cité\\
  INRIA\\
  INSERM\\
  F-75015, Paris, France\\

  Alexis Van Straaten\\
  HeKA\\
  Université Paris Cité\\
  INRIA\\
  INSERM\\
  F-75015, Paris, France\\
  \emph{and}\\
  Assistance Publique-Hôpitaux de Paris (AP-HP)\\
  Service d’informatique Médicale, Biostatistiques Et Santé Publique\\
  Hôpital Européen Georges Pompidou\\
  75015 Paris, France\\

  Emmanuel Bacry\\
  CEntre de REcherche en MAthématiques de la DÉcision\\
  CNRS-UMR 7534\\
  Université Paris-Dauphine PSL\\
  75016 Paris, France\\

  Anne-Sophie Jannot\\
  HeKA\\
  Université Paris Cité\\
  INRIA\\
  INSERM\\
  F-75015, Paris, France\\
  \emph{and}\\
  Banque Nationale de Données Maladies Rares,\\
  Direction des Services Numériques \\
  AP-HP\\
  Paris, France\\

  Jean Feydy\\
  HeKA\\
  Université Paris Cité\\
  INRIA\\
  INSERM\\
  F-75015, Paris, France\\

}

\begin{document}
\raggedbottom
\newpage


\section[Introduction]{Introduction} \label{sec:intro}

\paragraph{Scalable survival analysis.}
Survival analysis is used to analyze time-to-event data. The Cox proportional hazards model \citep{cox_regression_1972} is one of the most widely used models for survival analysis and can be extended to handle time-dependent covariates. The weighted cumulative exposure (WCE) model \citep{sylvestre_flexible_2009} extends this framework to time-dependent exposures by summarizing past exposures through a time-lagged weight function that captures the cumulative impact of historical exposure. This approach is particularly useful in longitudinal pharmacoepidemiological research \citep{pazzagli_methods_2018, sapirpichhadze_immunosuppression_2024, dayan_weighted_2025}.

Despite their widespread utility, traditional implementations of the Cox model and its extensions face substantial barriers when applied to modern, large-scale databases. These computational bottlenecks arise from two primary sources:

\begin{itemize}
    \item \textbf{Algorithmic complexity.} Methods incorporating time-dependent covariates typically rely on a counting-process representation, in which each individual contributes one or more rows corresponding to intervals where covariate values remain unchanged. If $n$ denotes the total number of such rows, repeated risk-set evaluations may scale as $\mathcal{O}(n^2)$ in standard implementations \citep{yang_massive_2024}. Consequently, cohorts with hundreds of thousands of individuals can yield even larger numbers of follow-up intervals, making serial CPU implementations prohibitively time-consuming.

    \item \textbf{Iterative computational demands.} To address high-dimensional covariate spaces, tune model specifications, and quantify uncertainty, practitioners increasingly rely on regularization, cross-validation, and resampling-based procedures such as bootstrapping \citep{efron_bootstrap_1979}. Interpretation techniques such as Shapley values \citep{sundararajan2020many} can also require many repeated model evaluations. These procedures multiply the computational burden by several orders of magnitude.
\end{itemize}

\paragraph{Existing packages.}
Survival analysis software in \proglang{R} and \proglang{Python} has traditionally relied on CPU-based execution. In \proglang{R}, the \pkg{survival} package \citep{survival_package} serves as the de facto standard, providing the foundation for domain-specific extensions such as the \pkg{WCE} package \citep{wce_package}, which implements the weighted cumulative exposure model. The \proglang{Python} ecosystem offers robust alternatives, including \pkg{scikit-survival} \citep{sksurv}, \pkg{lifelines} \citep{lifelines}, \pkg{statsmodels} \citep{seabold2010statsmodels}, and \pkg{tick} \citep{bacry2018tick}. While some of these libraries are highly optimized, they are primarily CPU-oriented and do not directly target GPU-accelerated fitting of Cox and WCE models.

Recent efforts to leverage massively parallel graphics processing units (GPUs) for survival analysis have followed two distinct architectural trajectories:

\begin{itemize}
    \item \textbf{Low-level \proglang{CUDA}/\proglang{C++} implementations.} Optimized implementations such as \pkg{CUDA\textunderscore Cox} \citep{andalibi2016effective}, \pkg{xgboost} \citep{barnwal2022survival}, and \pkg{Cyclops} \citep{yang_massive_2024} use compiled backends or specialized optimization routines to improve scalability. While these tools can achieve high performance, their rigid architecture may make it difficult for researchers to modify the underlying objective functions or incorporate custom penalties.

    \item \textbf{Deep learning frameworks.} Libraries such as \pkg{pycox} \citep{pycox} and \pkg{TorchSurv} \citep{monod_torchsurv_2024} use the high-level \pkg{PyTorch} framework \citep{paszke2017automatic}. They trade some degree of low-level optimization for the flexibility of automatic differentiation and seamless integration into modern machine learning pipelines.
\end{itemize}

In this context, there remains a need for a solution that bridges these two worlds: a package that provides the computational efficiency of an optimized GPU backend while maintaining the familiar syntax and statistical rigor of reference packages such as \proglang{R}'s \pkg{survival}.

\paragraph{Contributions.}
To address this challenge, we introduce \pkg{survivalGPU}, a high-performance package for large-scale survival analysis with the following key features:

\begin{itemize}
    \item \textbf{Model support.} We implement Cox and WCE models with native support for bootstrap estimation.

    \item \textbf{Accessible interfaces.} We provide user-friendly interfaces for both \proglang{R} and \proglang{Python}.

    \item \textbf{Modular and efficient GPU backend.} We use \pkg{PyTorch} for automatic differentiation and tensor-based optimization, and \pkg{KeOps} \citep{Keops} for memory-efficient estimation of the WCE-derived time-dependent covariates. This architecture provides substantial computational gains while remaining extensible through pure \proglang{Python}.
    
    \item \textbf{Reliability.} We show numerical agreement with reference implementations within precision-specific relative tolerances in both \code{float32} and \code{float64}.

   \item \textbf{Performance.} Benchmark experiments demonstrate substantial speedups. For Cox models, at \(n = 1{,}000{,}000\), \code{coxphGPU()} required 2.75 seconds on the GPU, compared with 15.35 seconds for \code{coxph()} from the \pkg{survival} package, corresponding to a \(5.6\times\) speedup. With \(B = 1{,}000\) bootstrap replicates on \(n = 100{,}000\) patients, \code{coxphGPU()} required 71.64 seconds (1.19 minutes), compared with 1{,}004.62 seconds (16.74 minutes) for the extrapolated \pkg{survival} runtime, corresponding to a \(14.0\times\) speedup. Gains were larger for WCE models: at \(n = 100{,}000\), \code{wceGPU()} required 9.83 seconds, compared with 9{,}050.72 seconds (2.51 hours) for \code{WCE()} from the \pkg{WCE} package, corresponding to a \(920.7\times\) speedup. With \(B = 1{,}000\) bootstrap replicates, \code{wceGPU()} required 5{,}021.78 seconds (1.39 hours), compared with an extrapolated runtime of 6{,}635{,}933.69 seconds (76.80 days) for \code{WCE()}, corresponding to a \(1{,}321.4\times\) speedup.
\end{itemize}

Our work is motivated by the emergence of nation-scale medico-administrative databases, such as the Système National des Données de Santé (SNDS) in France, which have transformed clinical research by enabling longitudinal studies on millions of patients \citep{maillard_use_2024, cairat_ingestion_2025, wyplosz_pneumococcal_2022, soeiro_high-dose_2025}.

The remainder of the paper is organized as follows. Section~\ref{sec:models} introduces the survival analysis framework, with a focus on the Cox proportional hazards model and its WCE extension. Section~\ref{sec:gpu} describes the implementation of \pkg{survivalGPU}, including the use of \pkg{torch} and \pkg{KeOps} for accelerated computation. Section~\ref{sec:tuto} presents the \proglang{Python} and \proglang{R} interfaces. Section~\ref{sec:validation} reports the implementation validation in both \code{float32} and \code{float64}, and Section~\ref{sec:benchmark} presents the benchmark study against existing reference implementations using \code{float32}. Section~\ref{sec:empirical} provides an empirical illustration of the package. Finally, Section~\ref{sec:conclusion} concludes with a discussion of current limitations and future developments.


\section{Survival analysis} \label{sec:models}

\subsection{Time-to-event data} \label{subsec:censored_data}

Survival analysis models time-to-event data. An event is an outcome of interest, such as the death of a patient or the diagnosis of a disease \citep{clark_survival_2003}. Event times may be censored, meaning that the exact event time is not fully observed within an individual-specific observation window. With right censoring, the event time is known only to exceed the censoring time. With left censoring, the event time is known only to precede the censoring time.

\paragraph{Notation.}
We define \(T \geq 0\) as a random variable representing a survival or event time. We denote by \(S(\cdot)\) the survival function, which represents the probability that an individual remains event-free past time \(t\):
\begin{equation}
S(t) = \mathbb{P}(T > t)~.
\label{eq:the_survival_function}
\end{equation}

The hazard function \(h(\cdot)\) is the instantaneous rate of occurrence of events among individuals who are still at risk at time \(t\):
\begin{equation}
h(t) =  \lim_{\Delta t \to 0^+} \ \frac{\mathbb{P}(t \leq T < t + \Delta t \mid T \geq t)}{\Delta t}~.
\label{eq:the_hazard_function}
\end{equation}

\subsection{The Cox model} \label{subsec:cox}

The Cox proportional hazards model \citep{cox_regression_1972} is one of the most widely used models in survival analysis. For a model with \(q\) covariates, possibly time-dependent but with time-fixed coefficients, the hazard function for individual \(j\) at time \(t\) is
\begin{equation}
h_j(t \mid \mathbf{X}_j(t)) = h_0(t)
\exp\left( \boldsymbol{\beta}^\top \mathbf{X}_j(t) \right)
=
h_0(t)
\exp\left( \sum_{s=1}^{q} \beta_s X_{js}(t) \right),
\label{eq:hazard_cox_time_dependent_multivariate}
\end{equation}
where \(h_0(t)\) is the unspecified baseline hazard, \(\mathbf{X}_j(t) = \left( X_{j1}(t), \ldots, X_{jq}(t) \right)\) is the vector of covariate values for individual \(j\), and \(\boldsymbol{\beta} = \left( \beta_1, \ldots, \beta_q \right)\) is the vector of regression coefficients.

The model is semi-parametric: the effect of the covariates is modeled through the exponential of the linear predictor \(\boldsymbol{\beta}^\top \mathbf{X}_j(t)\), while the shape of the baseline hazard function is left unspecified. The proportional hazards assumption implies that, for two individuals with covariate vectors \(\mathbf{X}_a(t)\) and \(\mathbf{X}_b(t)\), the hazard ratio is
\begin{equation}
\frac{h_a(t \mid \mathbf{X}_a(t))}{h_b(t \mid \mathbf{X}_b(t))}
=
\exp\left\{
\boldsymbol{\beta}^\top
\left(\mathbf{X}_a(t) - \mathbf{X}_b(t)\right)
\right\}.
\end{equation}
For time-independent covariates, this ratio is constant over time. In particular, a one-unit increase in covariate \(X_s\), all other covariates being equal, multiplies the hazard by \(e^{\beta_s}\).

The regression coefficients are estimated by maximizing the Cox partial likelihood. Let \(t_i\), \(i = 1,\ldots,k\), denote the ordered event times, and let \(r_i\) denote the individual experiencing the event at time \(t_i\). Assuming distinct event times, the partial likelihood is
\begin{equation}
PL(\boldsymbol{\beta}) =
\prod_{i=1}^{k}
\frac{
\exp\left( \boldsymbol{\beta}^\top \mathbf{X}_{r_i}(t_i) \right)
}{
\sum_{j \in R(t_i)}
\exp\left( \boldsymbol{\beta}^\top \mathbf{X}_j(t_i) \right)
}~,
\label{partial_likelihood}
\end{equation}
where \(k\) is the number of observed events and \(R(t_i)\) is the risk set at time \(t_i\), defined as the set of individuals still under observation just before \(t_i\). Tied event times require specific approximations, such as the Breslow or Efron methods. Standard likelihood-based optimization and inference procedures can then be applied to the partial likelihood to estimate the coefficient vector \(\boldsymbol{\beta}\).

\subsection{The WCE model} \label{subsec:wce_model}

The weighted cumulative exposure (WCE) model \citep{sylvestre_flexible_2009} extends the Cox model by replacing the current exposure value with a weighted summary of past exposures. Consider a cohort of patients in which exposure intensity varies over time. At any follow-up time \(u\), the exposure history is summarized by the WCE metric
\begin{equation}
 \mathrm{WCE}(u) = \sum_{t=0}^{u} w(u - t)\, X(t),
\label{wce_def}
\end{equation}
where \(t \leq u\) indexes previous exposure times, \(X(t)\) is the exposure at time \(t\), and \(w(u-t)\) is the weight assigned according to the time elapsed since exposure. The WCE metric \(\mathrm{WCE}(u)\) is then used as a time-dependent covariate in the Cox model:
\begin{equation}
h(u \mid \mathbf{X}(u), \mathbf{Z}(u)) =
h_0(u) \exp\left(
\beta \sum_{t=0}^{u} w(u - t) X(t)
+ \sum_{s=1}^{q} \eta_s Z_s(u)
\right),
\label{wce_in_cox}
\end{equation}
where \(\mathbf{X}(u) = \{X(t): 0 \leq t \leq u\}\) is the exposure history up to time \(u\), \(\mathbf{Z}(u) = \left(Z_1(u), \ldots, Z_q(u) \right)\) is the vector of additional Cox covariates, and \(\boldsymbol{\eta} = (\eta_1, \ldots, \eta_q)\) is the vector of their coefficients.

In practice, \(w(\tau)\) is set to zero for lags \(\tau > a\), where \(a\) is the cutoff defining the maximum lag over which past exposures are assumed to influence the current hazard.

The weight function \(w(\cdot)\) is modeled using a linear combination of cubic B-splines:
\begin{equation}
w(u - t) = \sum_{j=1}^{m} \theta_j B_j(u - t),
\label{bspline}
\end{equation}
where \(B_j(\cdot)\), \(j = 1, \ldots, m\), are the spline basis functions and \(\theta_j\) are their coefficients. Substituting equation \eqref{bspline} into equation \eqref{wce_in_cox} gives
\begin{equation}
h(u \mid \mathbf{X}(u), \mathbf{Z}(u)) =
h_0(u) \exp \left(
\sum_{j=1}^{m} \beta \theta_j
\sum_{t=0}^{u} B_j(u - t) X(t)
+ \sum_{s=1}^{q} \eta_s Z_s(u)
\right).
\label{wce_spline_cox}
\end{equation}

Conceptually, \(w(\cdot)\) describes the shape of the weight function and \(\beta\) the magnitude of the cumulative exposure effect. However, both \(\beta\) and the spline coefficients \(\theta_j\) determine the strength of the association between exposure and risk, which creates an identifiability problem. Following \citet{sylvestre_flexible_2009}, we define
\begin{equation}
\gamma_j = \beta \theta_j, \quad \text{for~~} j = 1, \ldots, m.
\end{equation}

Estimation is then simplified by defining artificial time-dependent covariates, called WCE-derived covariates,
\begin{equation}
D_j(u) = \sum_{t=0}^{u} B_j(u - t) X(t),
\quad\text{for~~} j = 1, \ldots, m.
\label{wce_feature}
\end{equation}

The hazard function can therefore be written in the familiar Cox model form:
\begin{equation}
h(u \mid \mathbf{X}(u), \mathbf{Z}(u)) =
h_0(u) \exp\left(
\sum_{j=1}^{m} \gamma_j D_j(u)
+ \sum_{s=1}^{q} \eta_s Z_s(u)
\right),
\label{wce_cox_final}
\end{equation}
where \(\mathbf{D}(u) = (D_1(u), \ldots, D_m(u))\) is the vector of WCE-derived covariates, and \(\boldsymbol{\gamma} = (\gamma_1, \ldots, \gamma_m)\) is the vector of their estimable coefficients.

\paragraph{Estimation.}
Once the spline basis has been specified, the WCE-derived covariates \(D_j(u)\) can be computed for each follow-up time. The Cox partial likelihood in equation \eqref{partial_likelihood} can then be maximized to estimate \(\boldsymbol{\gamma}\) and \(\boldsymbol{\eta}\).

\paragraph{Hazard ratios.}
Because \(\beta\) is absorbed into \(\boldsymbol{\gamma}\), the fitted model does not separately estimate a scalar exposure coefficient and a normalized weight function. Instead, hazard ratios are obtained by comparing exposure histories. For two hypothetical exposure histories \(\mathbf{X}_1\) and \(\mathbf{X}_0\), evaluated at time \(u\) with identical values of the adjustment covariates, the estimated hazard ratio is
\begin{equation}
\widehat{\mathrm{HR}}(\mathbf{X}_1, \mathbf{X}_0; u) =
\exp\left(
\sum_{j=1}^{m} \hat{\gamma}_j
\sum_{t=0}^{u} B_j(u - t)
\left[X_1(t) - X_0(t)\right]
\right).
\label{eq:hr_wce}
\end{equation}

Confidence intervals for such contrasts can be obtained using patient-level bootstrap resampling \citep{efron_bootstrap_1979}.

\paragraph{Hyperparameters.}
The previous equations assume that the spline basis has already been fixed. In practice, this basis depends on hyperparameters that control the flexibility and support of the estimated weight function.

\begin{itemize}
    \item \textbf{\code{nknots}}: the number of interior knots used in the cubic B-spline representation of the weight function. A cubic B-spline basis with \(m\) basis functions corresponds to \(m - 4\) interior knots. Increasing the number of knots increases the flexibility of the estimated weight function, but also increases the risk of overfitting. In practice, following \citet{sylvestre_flexible_2009}, a small number of interior knots, typically between 1 and 3, is usually sufficient.

    \item \textbf{\code{cutoff}}: the maximum lag \(a\) over which past exposures are assumed to influence the current hazard. The weight function \(w(\cdot)\) is estimated on the interval \([0,a]\), and exposures occurring before \(u-a\) are assigned a weight of zero. The cutoff therefore defines the time window of etiologically relevant exposure. It also determines the scale over which the spline knots are distributed: a large cutoff spreads the knots over a wider interval, whereas a small cutoff concentrates them on more recent exposure times.

    \item \textbf{\code{constrained}}: whether a boundary constraint is imposed at the right end of the exposure window. In the right-constrained specification, the last two spline coefficients are fixed to zero so that the estimated weight function reaches zero smoothly at the cutoff. This constraint is useful when it is assumed that exposures older than the cutoff should have no residual effect on the hazard.
\end{itemize}

\paragraph{Model selection.}
In practice, the model is fitted over a grid of hyperparameter values. In the analyses reported below, we consider 1 to 3 interior knots and both constrained and unconstrained specifications. Model selection is performed using the Bayesian information criterion (BIC) \citep{schwarz_estimating_1978}:
\begin{equation}
\mathrm{BIC} = -2 \ln(PL) + p \ln(d),
\label{eq:BIC}
\end{equation}
where \(PL\) denotes the partial likelihood, \(p\) is the number of estimated regression parameters, including the WCE-derived covariate coefficients and any adjustment covariate coefficients, and \(d\) is the number of observed events. The selected model is the specification with the lowest BIC.

The comparison between constrained and unconstrained specifications is informative for the choice of cutoff. If an unconstrained model assigns substantial weight near the right boundary, this may indicate that the cutoff is too short to capture the full exposure-risk relationship. Conversely, if the right-constrained model is selected, this is consistent with the assumption that the exposure effect vanishes by the end of the chosen time window.


\section{GPU-powered survival analysis} \label{sec:gpu}

\subsection[Architecture of the survivalGPU package]{Architecture of the \pkg{survivalGPU} package} \label{subsec:gpu_wce}

\paragraph{Package architecture.}
The \pkg{survivalGPU} package provides a GPU-accelerated implementation of Cox and WCE models. It is composed of a computational backend implemented in \proglang{Python} and an \proglang{R} interface that allows users to call the main functions within an \proglang{R} environment. Users can fit Cox and WCE models from either language, while the core computations are performed in \proglang{Python}, including GPU-specific routines.

\paragraph{Floating-point precision.}
The computational backend supports both single- and double-precision floating-point arithmetic, through \code{float32} and \code{float64} tensors, respectively. The choice of precision is important for GPU-based statistical software, because single precision requires less memory and is often substantially faster, whereas double precision may be preferred when numerical accuracy is the main concern. Similar trade-offs between single- and double-precision arithmetic have been discussed in previous GPU-oriented statistical software papers \citep{casarin_parallel_2015}. In \pkg{survivalGPU}, \code{float32} is therefore the recommended setting for GPU acceleration, especially on hardware with limited double-precision throughput, while \code{float64} remains available for sensitivity analyses. This precision option applies to both main computational steps described below: the construction of the WCE-derived covariates in Section~\ref{subsec:wce_model} and the maximization of the Cox partial likelihood in Section~\ref{subsec:cox}. In Section~\ref{sec:validation}, we compare the precision modes to assess numerical agreement with the reference implementations. In contrast, the computational benchmarks in Section~\ref{sec:benchmark} are reported using \code{float32}, corresponding to the recommended high-performance configuration.

\paragraph{Computational bottlenecks.}
From a computational standpoint, the main bottlenecks depend on the model being fitted. For the WCE model, the most intensive step is the construction of the artificial time-dependent covariates, referred to here as WCE-derived covariates, defined in equation~\eqref{wce_feature}. These covariates encode the exposure history of each individual through the spline basis and must be computed before maximizing the Cox partial likelihood.

For both Cox and WCE models, maximizing the Cox partial likelihood in equation~\eqref{partial_likelihood} also contributes to runtime. In the Cox model, this is the main computational step. In the WCE model, it follows the construction of the WCE-derived covariates and remains an important contributor to runtime.

\subsection{Estimation of the WCE-derived covariates} \label{subsec:gpu_covariates}

\paragraph{Construction of WCE-derived covariates.}
As described in Section~\ref{subsec:wce_model}, the WCE model transforms the exposure history \(X(t)\) into artificial time-dependent covariates by convolving it with the cubic B-spline basis functions \(B_j(\cdot)\). For each basis spline \(B_j\), the corresponding WCE-derived covariate at time \(u\) is
\begin{equation}
D_j(u) = \sum_{t=0}^{u} B_j(u - t)\, X(t).
\label{eq:gpu_wce_feature}
\end{equation}
In practice, terms with \(u-t>a\) are omitted because \(w(u-t)=0\) beyond the cutoff.

This operation must be repeated for each basis spline, each patient, and each follow-up time at which the time-dependent covariates are evaluated. It is therefore both computationally intensive and memory-demanding, especially when the cohort is large or when follow-up histories are long.

For a single individual, equation~\eqref{eq:gpu_wce_feature} can be written as a matrix-vector product. Let \(\mathbf{X}\) denote the vector of exposures over follow-up and let \(\mathbf{B}_j\) be the structured matrix whose \((u,t)\) entry is
\begin{equation}
(\mathbf{B}_j)_{ut} =
\begin{cases}
B_j(u-t), & t \leq u, \\
0, & t > u.
\end{cases}
\end{equation}
When a finite cutoff is used, entries with \(u-t>a\) are also zero, so the matrix is banded lower triangular.

Then the vector of values of the \(j\)th WCE-derived covariate can be written as
\begin{equation}
\mathbf{D}_j = \mathbf{B}_j \mathbf{X}.
\end{equation}

When several spline basis functions are used, this operation is repeated for \(j = 1,\ldots,m\), producing, at each follow-up time \(u\), the vector of WCE-derived covariates \(\mathbf{D}(u) = (D_1(u), \ldots, D_m(u))\) used in equation~\eqref{wce_cox_final}.

\paragraph{Sparse exposure histories.}
In many pharmacoepidemiological applications, exposure histories are sparse: patients are unexposed for long periods, and only time points with \(X(t) \neq 0\) and \(u-t \leq a\) contribute to the WCE-derived covariates. The implementation therefore stores and processes only nonzero exposure times. This sparse representation reduces both memory use and computation time, especially when drug exposure is intermittent or when follow-up is long.

\paragraph{Symbolic matrix products with \pkg{KeOps}.}
Standard tensor libraries, such as \pkg{PyTorch}, are optimized for dense, fixed-size tensor operations. By contrast, the construction of WCE-derived covariates involves irregular patient-specific exposure histories and large structured matrices that should not be explicitly materialized in memory. To address this, \pkg{survivalGPU} relies on \pkg{KeOps} \citep{Keops, feydy2020fast}, which supports symbolic matrix computations and lazy evaluation.

With \pkg{KeOps}, the expression defining the entries \(B_j(u-t)\) is represented symbolically and evaluated only when needed. This avoids storing the full matrices \(\mathbf{B}_j\), whose size increases quadratically with the number of follow-up times for each individual. The same symbolic formulation can be executed on the CPU, reducing memory allocation compared with explicit matrix construction, or on the GPU, where the computations are additionally parallelized.

\begin{figure}[H]
\centering
\includegraphics[width=0.75\textwidth]{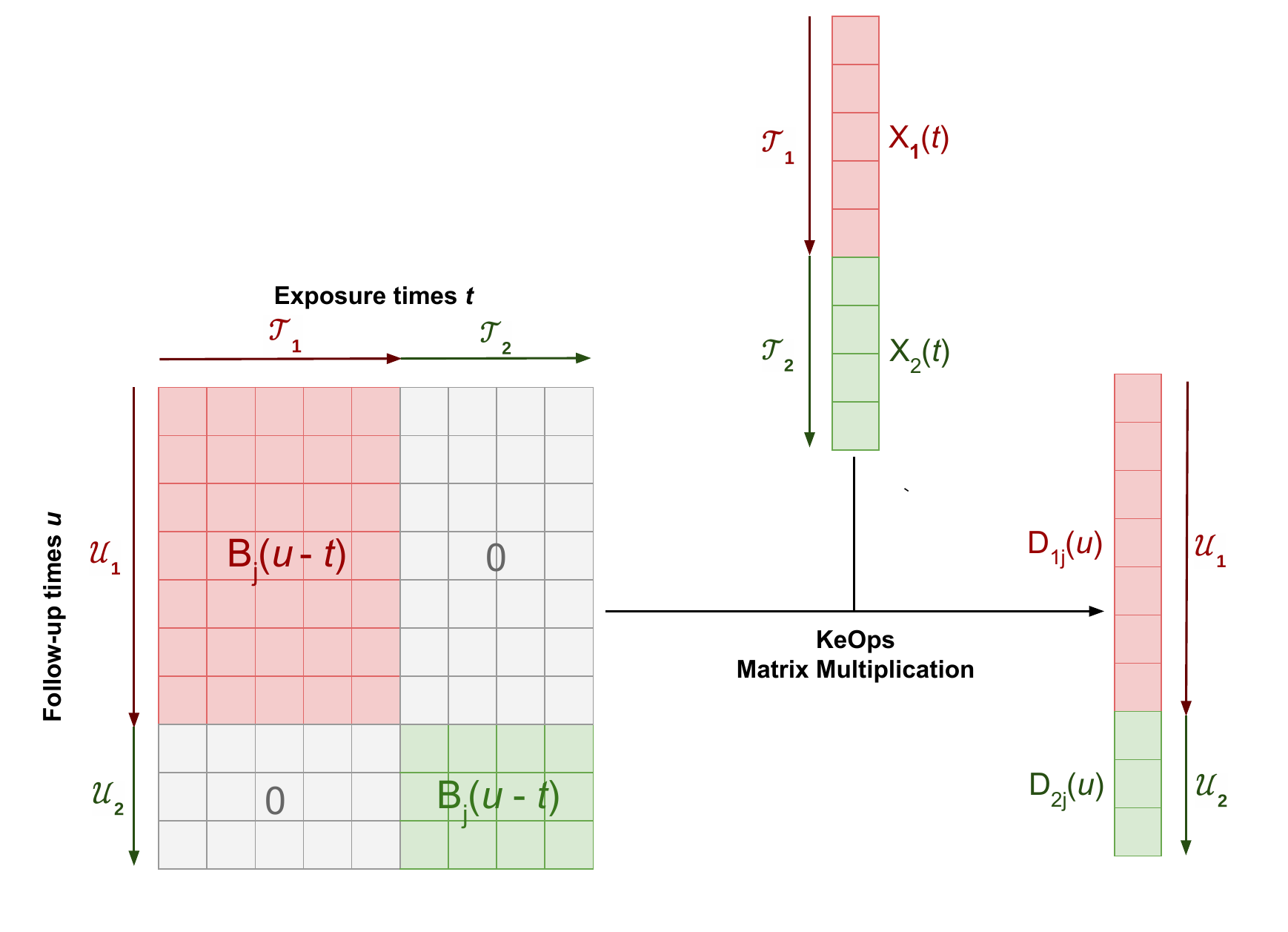}
\caption{Computation of the WCE-derived covariate \(D_j\) for one B-spline \(B_j\) and two patients. \(\mathcal{T}_i\) denotes the set of times \(t\) at which patient \(i\) receives a nonzero exposure \(X_i(t)\), and \(\mathcal{U}_i\) denotes the set of follow-up times \(u\) at which the WCE-derived covariate is evaluated. The matrix entries correspond to the spline weights \(B_j(u-t)\), and multiplication by the exposure vector yields the WCE-derived covariate values \(D_j(u)\). The use of \pkg{KeOps} avoids explicitly materializing the full matrix and excludes entries that do not contribute to the computation.}
\label{fig:wce_keops}
\end{figure}

Conceptually, the patient-specific matrices form the diagonal blocks of a larger block matrix. Off-diagonal blocks correspond to combinations of exposure and follow-up times from different patients and must not contribute to the WCE-derived covariates. The off-diagonal interactions are masked in the symbolic expression and are not materialized in memory. This block-diagonal structure allows \pkg{survivalGPU} to compute WCE-derived covariates for all patients using a single symbolic matrix product while preserving patient-specific exposure histories.

\clearpage

\subsection{Maximization of the Cox partial likelihood} \label{subsec:gpu_likelihood}

\paragraph{Optimization strategy.}
After the covariate matrix has been constructed, the Cox and WCE models are fitted by maximizing the Cox partial likelihood in equation~\eqref{partial_likelihood}. The implementation is conceptually inspired by the \pkg{survival} package in \proglang{R} \citep{survival_package}. However, whereas \pkg{survival} relies on compiled \proglang{C} routines, \pkg{survivalGPU} implements the optimization in \proglang{Python}, with tensor operations executed on the CPU or GPU depending on the selected device.

\paragraph{Automatic differentiation.}
In the \pkg{survival} package, optimization relies on compiled \proglang{C} routines with analytically derived derivatives. In \pkg{survivalGPU}, the derivatives required by the optimizer are computed using automatic differentiation in \pkg{PyTorch}. This avoids manually coding model-specific derivatives and makes the implementation easier to extend to alternative Cox or WCE model formulations, such as penalized objectives.

\paragraph{Extensibility.}
This architecture separates the definition of the objective function from the implementation of derivative calculations. As a result, extensions of the Cox or WCE likelihood can be implemented by modifying the objective function while relying on automatic differentiation for optimization. This design is useful for future extensions such as penalized estimation or additional Cox model structures.


\section{Package overview} \label{sec:tuto}

The \pkg{survivalGPU} package provides a GPU-accelerated implementation of Cox and WCE models. It is implemented in \proglang{Python} using the \code{torch} module of \pkg{PyTorch} and the \pkg{KeOps} package, as described in Section~\ref{sec:gpu}. The package can be used either directly from \proglang{Python} or through an \proglang{R} interface. The \proglang{R} interface relies on the \pkg{reticulate} package to call the \proglang{Python} backend, while providing a syntax close to standard \proglang{R} survival analysis packages. Internally, the \code{coxphGPU()} and \code{wceGPU()} functions call the \code{coxph_R()} and \code{wce_R()} \proglang{Python} functions, which take a \code{pandas.DataFrame} and column names as arguments and dispatch to the object-based estimators described below. This allows us to maintain the numerical part of the code in a single place, in \proglang{Python}. This makes the code simpler to develop and maintain, and reduces the risk of divergence between the two implementations, while making the package available in both \proglang{R} and \proglang{Python}.

Both interfaces expose the floating-point precision option described in Section~\ref{sec:gpu}. The backend can use either \code{float32} or \code{float64} arithmetic. The examples below use \code{float32}, which is the recommended configuration for GPU acceleration and the configuration used in the benchmark experiments reported in Section~\ref{sec:benchmark}.

Using the WCE model requires the \pkg{PyKeOps} package, which is available on Linux and macOS but not on Windows. Installing \pkg{PyKeOps} is optional: the Cox model can be used without it. On Windows, \pkg{PyKeOps} can be used through the Windows Subsystem for Linux (WSL). GPU acceleration uses \proglang{CUDA} through the \pkg{torch} and \pkg{PyKeOps} packages and requires an NVIDIA GPU. Code examples can be found in Section~\ref{sec:empirical}.

\subsection[Python package overview]{\proglang{Python} package overview} \label{subsec:python_tuto}

There exist multiple implementations of the Cox model in \proglang{Python}, such as the \pkg{lifelines} and \pkg{scikit-survival} packages. However, there is no single reference implementation. We decided to follow the conventions of the \pkg{scikit-learn} package whenever possible, as it is well known to \proglang{Python} users. Statistical models are implemented as estimator objects: users initialize an estimator with model-specific options and then estimate the model parameters using the \code{fit()} method, which stores the results as attributes of the object. Unlike \pkg{scikit-survival}, \code{fit()} takes \pkg{NumPy} arrays rather than a \code{DataFrame} with column names, because the low-level interface is designed to be called on data that are already loaded on the target device.

\paragraph{Cox model.}
The Cox proportional hazards model is implemented by the \code{CoxPHSurvivalAnalysis} estimator. Its arguments, and those of its \code{fit()} method, are listed in Table~\ref{tab:inputs-cox} alongside their equivalents in the \proglang{R} interface and in \code{coxph()} from \pkg{survival}, and its features are compared with existing implementations in Table~\ref{tab:comparison-cox}. Both the Breslow and Efron approximations for handling tied event times are implemented and can be selected through the \code{ties} argument. The number of bootstrap replicates and the batch size used to process them in parallel on the GPU are specified when the estimator is initialized, through the \code{nbootstraps} and \code{batchsize} arguments. Numerical precision is controlled by the \code{dtype} argument, which accepts \code{np.float32} or \code{np.float64}. Single precision reduces memory use for floating-point tensors and can improve runtime on GPU devices, whereas double precision may be preferred when numerical precision is the main concern.

The fitted object stores the estimated coefficients, their standard errors, and the model log-likelihood values as attributes (\code{coef_}, \code{std_}, \code{loglik_}, \ldots{}). If \code{nbootstraps} is positive, the bootstrap replicates of the coefficients are stored in \code{bootstrap_coef_}.

\begin{table}[t!]
\centering
\footnotesize
\setlength{\tabcolsep}{4pt}
\renewcommand{\arraystretch}{1.1}
\begin{tabular}{@{}>{\raggedright\arraybackslash}p{3.35cm}>{\raggedright\arraybackslash}p{3.65cm}>{\raggedright\arraybackslash}p{3.65cm}>{\raggedright\arraybackslash}p{3.65cm}@{}}
\toprule
 & \multicolumn{2}{c}{\pkg{survivalGPU}} & \pkg{WCE} \\
\cmidrule(lr){2-3}
 & \proglang{Python} & \proglang{R} & \proglang{R} \\
\midrule
\multicolumn{4}{@{}l}{\textit{Data}} \\
\hspace{0.8em}Data set & Arrays in \code{fit()} & \code{data} & \code{data} \\
\hspace{0.8em}Subject identifier & \code{patient} & \code{id} & \code{id} \\
\hspace{0.8em}Event indicator & \code{event} & \code{event} & \code{event} \\
\hspace{0.8em}Interval bounds & \code{start}, \code{stop} & \code{start}, \code{stop} & \code{start}, \code{stop} \\
\hspace{0.8em}Exposure & \code{dose} & \code{expos} & \code{expos} \\
\hspace{0.8em}Covariates & \code{covariates} (array) & \code{covariates} (names) & \code{covariates} (names) \\
\hspace{0.8em}Strata & \code{strata} in \code{fit()} & \code{strata()} special in \code{formula} & \code{strata()} special in \code{formula} \\\addlinespace[5pt]
\multicolumn{4}{@{}l}{\textit{Weight function and model selection}} \\
\hspace{0.8em}Number of knots & \code{nknots} (integer or list) & \code{nknots} (vector) & \code{nknots} (vector) \\
\hspace{0.8em}Time window & \code{cutoff} & \code{cutoff} & \code{cutoff} \\
\hspace{0.8em}Constraint & \code{constrained}: \code{None}, \code{"left"}, \code{"right"} & \code{constrained}: \code{FALSE}, \code{"Left"}, \code{"Right"} & \code{constrained}: \code{FALSE}, \code{"Left"}, \code{"Right"} \\
\hspace{0.8em}Information criterion & \code{criterion}: \code{"aic"}, \code{"bic"} & \code{aic}: \code{TRUE}, \code{FALSE} & \code{aic}: \code{TRUE}, \code{FALSE} \\
\addlinespace[5pt]
\multicolumn{4}{@{}l}{\textit{Bootstrap}} \\
\hspace{0.8em}Number of replicates & \code{nbootstraps} & \code{R} in \code{bootstrap()} & -- \\
\hspace{0.8em}Replicates per batch & \code{batchsize} & \code{batchsize} in \code{bootstrap()} & -- \\
\hspace{0.8em}Confidence level & \code{level} in \code{HR()} & \code{level in \code{HR()}} & -- \\
\addlinespace[5pt]
\multicolumn{4}{@{}l}{\textit{Computation}} \\
\hspace{0.8em}Precision & \code{dtype} (\code{np.float32}, \code{np.float64}) & \code{double\_precision} (\code{FALSE}, \code{TRUE}) & -- \\
\hspace{0.8em}Device & \code{device} & \code{device} & -- \\
\bottomrule
\end{tabular}
\caption{\label{tab:inputs-wce} Inputs of the WCE model in \pkg{survivalGPU} and in \code{WCE()} from the \pkg{WCE} package. In \proglang{Python}, options are given to the \code{WCESurvivalAnalysis} constructor and data to its \code{fit()} method. In \proglang{R}, \code{wceGPU()} keeps the argument names of \code{WCE()}, and bootstrap options are given to \code{bootstrap()} once the model is fitted. In both packages, the data set is in long format, with one row per subject and time unit. $^{a}$Only \code{"Cox"} is implemented in \pkg{WCE}.}
\end{table}

\paragraph{WCE model.}
The WCE model is implemented by the \code{WCESurvivalAnalysis} estimator. Its arguments, and those of its \code{fit()} method, are listed in Table~\ref{tab:inputs-wce} alongside their equivalents in the \proglang{R} interface and in \code{WCE()} from the \pkg{WCE} package, and its features are compared with those of \code{WCE()} in Table~\ref{tab:comparison-wce}. In addition to the survival outcome and adjustment covariates, the user specifies the exposure variable and the hyperparameters of the spline basis: the number of knots (\code{nknots}), the length of the exposure-risk window (\code{cutoff}), and the boundary constraint (\code{constrained}, one of \code{"left"}, \code{"right"}, or \code{None}). When several candidate values are given for \code{nknots}, one model is fitted for each candidate and the model that minimizes the information criterion selected by the \code{criterion} argument (\code{"aic"} or \code{"bic"}) is retained. As in the reference \pkg{WCE} package, tied event times are handled using the Efron approximation, which is the default in the underlying Cox model. As for \code{CoxPHSurvivalAnalysis}, the number of bootstrap replicates and the batch size are specified at initialization through \code{nbootstraps} and \code{batchsize}, and the \code{dtype} argument selects between single- and double-precision arithmetic.

The fitted object exposes the estimated risk function of the selected model as \code{risk_function_}, the coefficients of the adjustment covariates as \code{coef_}, and the selected number of knots as \code{best_nknots_}; the results for all candidates remain available in the \code{*_grid_} attributes. When \code{nbootstraps} is positive, all candidates are fitted on the same bootstrap resamples and the number of knots is selected again in each replicate. The bootstrap replicates of the risk function, stored in \code{bootstrap_risk_functions_}, therefore account for the uncertainty in the choice of the number of knots, and the number of knots selected in each replicate is stored in \code{bootstrap_best_nknots_}. Finally, the \code{HR()} method computes the hazard ratio between two exposure profiles from the estimated risk function, together with a bootstrap percentile confidence interval when \code{nbootstraps} is positive.

\begin{table}[t!]
\centering
\footnotesize
\setlength{\tabcolsep}{4pt}
\renewcommand{\arraystretch}{1.1}
\begin{tabular}{@{}>{\raggedright\arraybackslash}p{3.35cm}>{\raggedright\arraybackslash}p{3.65cm}>{\raggedright\arraybackslash}p{3.65cm}>{\raggedright\arraybackslash}p{3.65cm}@{}}
\toprule
 & \multicolumn{2}{c}{\pkg{survivalGPU}} & \pkg{survival} \\
\cmidrule(lr){2-3}
 & \proglang{Python} & \proglang{R} & \proglang{R} \\
\midrule
\multicolumn{4}{@{}l}{\textit{Data and model}} \\
\hspace{0.8em}Survival outcome & \code{start}, \code{stop}, \code{event} in \code{fit()} & \code{Surv(start, stop, event)} in \code{formula} & \code{Surv(start, stop, event)} in \code{formula} \\
\hspace{0.8em}Covariates & \code{covariates} (array) in \code{fit()} & Right-hand side of \code{formula} & Right-hand side of \code{formula} \\
\hspace{0.8em}Strata & \code{strata} in \code{fit()} & \code{strata()} in \code{formula} & \code{strata()} in \code{formula} \\
\hspace{0.8em}Ties & \code{ties} & \code{ties} & \code{ties} \\
\hspace{0.8em}Initial values & \code{init} in \code{fit()} & \code{init} & \code{init} \\
\hspace{0.8em}Maximum iterations & \code{maxiter} & \code{iter.max} in \code{control} & \code{iter.max} in \code{control} \\
\hspace{0.8em}Store the data & -- & \code{X} (\code{FALSE}, \code{TRUE}) & \code{X} (\code{FALSE}, \code{TRUE}) \\
\addlinespace[5pt]
\multicolumn{4}{@{}l}{\textit{Bootstrap}} \\
\hspace{0.8em}Number of replicates & \code{nbootstraps} & \code{R} in \code{bootstrap()} & -- \\
\hspace{0.8em}Patient identifier & \code{patient} in \code{fit()} & \code{patient\_id} in \code{bootstrap()} & -- \\
\hspace{0.8em}Replicates per batch & \code{batchsize} & \code{batchsize} in \code{bootstrap()} & -- \\
\addlinespace[5pt]
\multicolumn{4}{@{}l}{\textit{Computation}} \\
\hspace{0.8em}Precision & \code{dtype} (\code{np.float32}, \code{np.float64}) & \code{double\_precision} (\code{FALSE}, \code{TRUE}) & -- \\
\hspace{0.8em}Device & \code{device} & \code{device} & -- \\
\bottomrule
\end{tabular}
\caption{\label{tab:inputs-cox} Inputs of the Cox model in \pkg{survivalGPU} and in \code{coxph()} from \pkg{survival}. In \proglang{Python}, options are given to the \code{CoxPHSurvivalAnalysis} constructor and data to its \code{fit()} method. In \proglang{R}, \code{coxphGPU()} takes the same formula interface as \code{coxph()}, and bootstrap options are given to \code{bootstrap()} once the model is fitted. }
\end{table}

\newpage

\subsection[R package overview]{\proglang{R} package overview} \label{subsec:R_tuto}

In \proglang{R}, the package syntax follows the conventions of the \pkg{survival} package and the \code{WCE()} function from the \pkg{WCE} package, while adding GPU-specific features such as parallelized bootstrap resampling, batch-wise control of GPU memory use, and explicit control of floating-point precision. The package provides two main functions, \code{coxphGPU()} for Cox models and \code{wceGPU()} for WCE models, and a \code{bootstrap()} generic function. Following the usual practice in \proglang{R} of separating model fitting from inference, a model is first fitted without bootstrap, and bootstrap replicates are added afterwards by calling \code{bootstrap()} on the fitted object. The fitted model is not transmitted back to the \proglang{Python} backend: \code{bootstrap()} sends the data again, and the backend fits the model once more together with the bootstrap replicates. Because this fit is processed in parallel with the replicates, its additional cost is negligible. In both functions, the \code{double_precision} argument selects between double-precision (\code{TRUE}, the default, \code{float64}) and single-precision (\code{FALSE}, \code{float32}) arithmetic, and the \code{device} argument selects the device on which computations are performed.

\paragraph{Cox model.}
The function \code{coxphGPU()} fits a Cox proportional hazards model. Its syntax is analogous to \code{coxph()} from the \pkg{survival} package (Table~\ref{tab:inputs-cox}), and both the Breslow and Efron approximations for tied event times can be selected using the \code{ties} argument. Since \code{coxph()} is a function rather than an S3 generic, \code{coxphGPU()} cannot inherit its handling of the inputs. Instead, \code{coxphGPU()} calls \code{survival::coxph()} with \code{iter.max = 0}, so that no Newton iteration is performed, and retrieves the response, the design matrix and the strata that it builds. Formula parsing, including the \code{strata()} special, is thus delegated to the \pkg{survival} package without duplicating its code, and only the preprocessed data are sent to the \proglang{Python} backend. Arguments that the backend does not support, such as \code{weights}, \code{offset()}, \code{tt()}, \code{cluster()} and multi-state models, stop with an error.

The function returns an object of class \code{c("coxphGPU", "coxph")}, which inherits from the class of \code{coxph()} objects. The \code{print()}, \code{summary()} and \code{coef()} methods are redefined for \code{coxphGPU} objects, and \code{summary()} additionally reports bootstrap percentile confidence intervals when bootstrap replicates are available. Other generic functions, such as \code{predict()}, \code{residuals()}, \code{update()} and \code{cox.zph()}, rely on the methods of the \pkg{survival} package, and functions from packages that work with \code{coxph} objects, such as \code{ggforest()} from \pkg{survminer}, can be applied directly.

Bootstrap confidence intervals are obtained by calling \code{bootstrap(fit, R, patient_id, data, batchsize)}, where \code{fit} is a model fitted with \code{x = TRUE}. The \code{patient_id} argument identifies the subject associated with each row of \code{data}, so that resampling is performed at the patient level rather than at the row level. The \code{batchsize} argument controls the number of bootstrap replicates processed in parallel; if memory errors occur, users can reduce \code{batchsize} to limit GPU memory use. The function returns a copy of the fitted object to which the bootstrap replicates of the coefficients are added.

\paragraph{WCE model.}
The function \code{wceGPU()} fits a WCE model within the Cox proportional hazards framework. It follows the nomenclature of the \code{WCE()} function from the \pkg{WCE} package (Table~\ref{tab:inputs-wce}) and uses WCE-specific hyperparameters such as \code{nknots}, \code{cutoff} and \code{constrained}. As in the \pkg{WCE} package, tied event times are handled using the Efron approximation. When several values are given for \code{nknots}, one model is fitted for each candidate and, unlike \code{WCE()}, which returns all fitted models and leaves the choice to the user, \code{wceGPU()} retains the model that minimizes the BIC, or the AIC when \code{aic = TRUE}. The selected number of knots is stored in \code{best.nknots}, and the information criterion and partial log-likelihood of every candidate are kept in the \code{*.grid} fields and displayed by \code{summary(allres = TRUE)}.

The function returns an object of class \code{"wceGPU"}. This class does not inherit from the class of \code{WCE()} objects: a \code{wceGPU} object describes the selected model only, whereas a \code{WCE} object contains every fitted model, so the methods of the \pkg{WCE} package would not apply. Instead, \code{wceGPU} objects have their own \code{print()}, \code{summary()}, \code{coef()}, \code{confint()} and \code{plot()} methods, whose output follows that of the \pkg{WCE} package, and an \code{HR()} function computes the hazard ratio between two exposure profiles.

Bootstrap confidence intervals are obtained by calling \code{bootstrap(fit, R, data, batchsize)}. All candidate numbers of knots are fitted on the same resamples, and the number of knots is selected again in each replicate, so that the confidence intervals account for the uncertainty in this choice; the number of knots selected in each replicate is stored in \code{bootstrap.best.nknots}. The \code{plot()} method then displays the estimated weight function together with its bootstrap confidence band, and \code{HR()} reports a bootstrap percentile confidence interval for the hazard ratio.


\section{Implementation validation} \label{sec:validation}

The objective of the implementation validation was to verify that \pkg{survivalGPU} reproduces results obtained with established reference implementations. This validation focuses on numerical differences between software implementations rather than on the statistical behavior of the estimators. Details on the data generation procedure and simulation scenarios are provided in Appendix~\ref{sec:simulation}. The benchmark study uses a separate simulation setting described in Section~\ref{sec:benchmark}.

As described in Section~\ref{sec:gpu}, \pkg{survivalGPU} supports both single- and double-precision floating-point arithmetic. The validation was therefore performed in both \code{float32} and \code{float64}. This comparison served two objectives. First, it verifies that the implementation reproduces the reference software results when using double precision. Second, it quantifies the numerical differences introduced by the recommended \code{float32} GPU configuration used in the benchmark experiments of Section~\ref{sec:benchmark}.

\paragraph{Cox model validation.}
The Cox implementation was validated in two steps. First, we adapted unit tests from the \pkg{survival} package to verify that \code{coxphGPU()} reproduces the behavior of \code{coxph()} for standard Cox model configurations, including right-censored and counting-process inputs, fixed and time-dependent covariates, tied event times, and the Breslow and Efron approximations. Second, simulation-based comparisons were used to quantify numerical differences between \code{coxphGPU()} and \code{coxph()} across repeated simulated data sets.

Four simulation settings were considered: one constant covariate, three constant covariates, one time-dependent covariate, and three time-dependent covariates. Each configuration was replicated \(N = 100\) times. In each replicate, the hazard ratio associated with each covariate was randomly selected from the set \(\{0.25, 0.5, 0.75, 1.5, 2.0, 3.0, 4.0\}\), and the corresponding regression coefficient was defined on the log-hazard-ratio scale as
\begin{equation}
\beta_s = \log(\mathrm{HR}_s),
\quad s = 1,\ldots,p,
\end{equation}
where \(p\) denotes the number of covariates in the simulation setting. This sampling scheme includes both protective effects, such as \(\log(0.25)\), and increased risks, such as \(\log(4.0)\). For each simulated data set, the same model was fitted with \code{coxphGPU()} and \code{coxph()}.

Both the Breslow and Efron tied event-time approximations were evaluated. For covariate \(s\) in replicate \(r\), the coefficient-level relative error was defined as
\begin{equation}
\mathrm{RE}_{rs} =
\frac{
\left|
\hat{\beta}^{\mathrm{GPU}}_{rs}
-
\hat{\beta}^{\mathrm{survival}}_{rs}
\right|
}{
\left|
\hat{\beta}^{\mathrm{survival}}_{rs}
\right|
}.
\label{eq:cox_relative_error_covariate}
\end{equation}
The replicate-level mean relative error was then defined as
\begin{equation}
\mathrm{MRE}_{r} =
\frac{1}{p}
\sum_{s=1}^{p}
\mathrm{RE}_{rs}.
\label{eq:cox_mre_replicate}
\end{equation}

When \(p = 1\), \(\mathrm{MRE}_{r}\) reduces to the relative error for the single coefficient. Table~\ref{tab:cox_validation_precision} reports the median and maximum of \(\mathrm{MRE}_{r}\) across the \(N = 100\) simulated data sets.

\begin{table}[t!]
\centering
\small
\setlength{\tabcolsep}{3.5pt}
\renewcommand{\arraystretch}{0.95}
\begin{tabular*}{\textwidth}{@{\extracolsep{\fill}}ccrrrr}
\hline
\noalign{\vskip 2pt}
& & \multicolumn{2}{c}{\code{float32}} & \multicolumn{2}{c}{\code{float64}} \\
\cline{3-4} \cline{5-6}
Scenario & Ties
& \multicolumn{1}{c}{Median MRE} & \multicolumn{1}{c}{Max MRE}
& \multicolumn{1}{c}{Median MRE} & \multicolumn{1}{c}{Max MRE} \\
\hline
1 const. cov.
& Efron
& \(4.11 \times 10^{-5}\) & \(5.89 \times 10^{-3}\)
& \(8.82 \times 10^{-11}\) & \(9.90 \times 10^{-7}\) \\

1 const. cov.
& Breslow
& \(5.14 \times 10^{-5}\) & \(7.50 \times 10^{-3}\)
& \(3.46 \times 10^{-13}\) & \(2.24 \times 10^{-6}\) \\

3 const. cov.
& Efron
& \(7.28 \times 10^{-5}\) & \(8.04 \times 10^{-4}\)
& \(1.08 \times 10^{-10}\) & \(3.17 \times 10^{-8}\) \\

3 const. cov.
& Breslow
& \(5.96 \times 10^{-5}\) & \(4.66 \times 10^{-4}\)
& \(1.24 \times 10^{-11}\) & \(1.75 \times 10^{-7}\) \\

1 time-dep. cov.
& Efron
& \(3.38 \times 10^{-5}\) & \(1.37 \times 10^{-3}\)
& \(1.22 \times 10^{-10}\) & \(1.25 \times 10^{-7}\) \\

1 time-dep. cov.
& Breslow
& \(3.14 \times 10^{-5}\) & \(1.36 \times 10^{-3}\)
& \(3.30 \times 10^{-11}\) & \(7.62 \times 10^{-8}\) \\

3 time-dep. cov.
& Efron
& \(5.05 \times 10^{-6}\) & \(5.75 \times 10^{-4}\)
& \(1.48 \times 10^{-10}\) & \(2.30 \times 10^{-8}\) \\

3 time-dep. cov.
& Breslow
& \(4.91 \times 10^{-6}\) & \(5.68 \times 10^{-4}\)
& \(7.92 \times 10^{-12}\) & \(1.01 \times 10^{-7}\) \\
\hline
\end{tabular*}
\caption{Cox model validation in \code{float32} and \code{float64}. Comparison of \code{coxphGPU()} from \pkg{survivalGPU} with \code{coxph()} from the \pkg{survival} package across \(N = 100\) simulated data sets of \(1{,}000\) patients. Results are reported separately for the Efron and Breslow tied event-time approximations. The scenario labels ``const. cov.'' and ``time-dep. cov.'' denote constant covariates and time-dependent covariates, respectively. For each replicate, coefficient-level relative errors were computed using Equation~\eqref{eq:cox_relative_error_covariate} and averaged across covariates to obtain \(\mathrm{MRE}_{r}\), as defined in Equation~\eqref{eq:cox_mre_replicate}. The table reports the median and maximum of \(\mathrm{MRE}_{r}\) across simulations for both numerical precisions.}
\label{tab:cox_validation_precision}
\end{table}

The Cox validation results show close numerical agreement with the reference implementation. In \code{float32}, median MRE values were below \(8 \times 10^{-5}\) in all scenarios, and the largest maximum MRE was \(7.50 \times 10^{-3}\). In \code{float64}, median MRE values were below \(2 \times 10^{-10}\), and the largest maximum MRE was \(2.24 \times 10^{-6}\). These results indicate that \code{coxphGPU()} reproduces \code{coxph()} closely in double precision, while retaining small numerical discrepancies in the recommended \code{float32} configuration.

\paragraph{WCE model validation.}
The WCE implementation was validated in two steps. First, unit tests compared \code{wceGPU()} from \pkg{survivalGPU} with the reference \code{WCE()} function from the \pkg{WCE} package on standard model specifications, including the construction of WCE-derived covariates, regression coefficients, variance estimates, partial log-likelihoods, information criteria, and hazard-ratio contrasts. Second, simulation-based comparisons were used to quantify numerical differences between \code{wceGPU()} and \code{WCE()} across randomly generated WCE scenarios.

For the simulation-based validation, each replicate combined a randomly selected weight-function scenario and a randomly selected target hazard ratio. The weight-function scenario was randomly selected from among the six non-null scenarios shown in Figure~\ref{fig:wce_shapes} of Appendix~\ref{sec:simulation}. The target hazard ratio was randomly selected from the set \(\{0.25, 0.5, 0.75, 1.5, 2.0, 3.0, 4.0\}\), and the corresponding WCE coefficient was defined on the log-hazard-ratio scale as
\begin{equation}
\beta_{\mathrm{WCE},r} = \log(\mathrm{HR}_r).
\end{equation}
Because the simulated weight functions are normalized over the 180-day lag window, this target hazard ratio can be interpreted as the hazard ratio comparing a patient receiving a constant unit dose during the full cutoff window with an otherwise identical unexposed patient. The cutoff was fixed at 180 days in all WCE validation analyses.

For each simulated data set, both unconstrained and right-constrained WCE specifications were fitted with \code{wceGPU()} and \code{WCE()} using 1, 2, and 3 interior knots. No model selection based on BIC was performed. Because the WCE model is estimated through spline coefficients rather than through a single exposure coefficient, validation focused on the standardized hazard-ratio contrast defined in Equation~\eqref{eq:hr_wce}. This contrast compared a hypothetical patient receiving a constant unit dose during the entire 180-day cutoff window with a hypothetical patient receiving a dose of zero during the same window.

For each replicate and fitted specification, we computed the HR estimated by \code{wceGPU()} and the HR estimated by the reference \code{WCE()} function. The relative error for the HR contrast was defined as
\begin{equation}
\mathrm{RE}_{r}(\mathrm{HR}) =
\frac{
\left|
\widehat{\mathrm{HR}}^{\mathrm{GPU}}_{r}
-
\widehat{\mathrm{HR}}^{\mathrm{WCE}}_{r}
\right|
}{
\left|
\widehat{\mathrm{HR}}^{\mathrm{WCE}}_{r}
\right|
}.
\label{eq:wce_re_hr}
\end{equation}
Table~\ref{tab:wce_validation_precision} reports the median and maximum of this relative error across the \(N = 100\) simulated data sets, separately for unconstrained and right-constrained WCE specifications and for each number of interior knots.

\begin{table}[t!]
\centering
\small
\setlength{\tabcolsep}{4pt}
\renewcommand{\arraystretch}{0.95}
\begin{tabular*}{\textwidth}{@{\extracolsep{\fill}}ccrrrr}
\hline
\noalign{\vskip 3pt}
& & \multicolumn{2}{c}{\code{float32}} & \multicolumn{2}{c}{\code{float64}} \\
\cline{3-4} \cline{5-6}
Interior knots & Constraint
& \multicolumn{1}{c}{Median RE} & \multicolumn{1}{c}{Max RE}
& \multicolumn{1}{c}{Median RE} & \multicolumn{1}{c}{Max RE} \\
\hline
\multirow{2}{*}{1}
& Unconstrained
& \(7.55 \times 10^{-5}\) & \(2.15 \times 10^{-3}\)
& \(5.36 \times 10^{-8}\) & \(8.47 \times 10^{-7}\) \\
& Right-constrained
& \(3.12 \times 10^{-5}\) & \(2.03 \times 10^{-3}\)
& \(4.27 \times 10^{-8}\) & \(8.06 \times 10^{-7}\) \\

\multirow{2}{*}{2}
& Unconstrained
& \(3.59 \times 10^{-5}\) & \(1.84 \times 10^{-3}\)
& \(3.98 \times 10^{-8}\) & \(1.36 \times 10^{-6}\) \\
& Right-constrained
& \(4.21 \times 10^{-5}\) & \(2.30 \times 10^{-3}\)
& \(4.05 \times 10^{-8}\) & \(3.34 \times 10^{-7}\) \\

\multirow{2}{*}{3}
& Unconstrained
& \(3.80 \times 10^{-5}\) & \(2.67 \times 10^{-3}\)
& \(5.28 \times 10^{-8}\) & \(2.39 \times 10^{-6}\) \\
& Right-constrained
& \(3.75 \times 10^{-5}\) & \(1.70 \times 10^{-3}\)
& \(5.35 \times 10^{-8}\) & \(9.40 \times 10^{-7}\) \\
\hline
\end{tabular*}
\caption{WCE model validation in \code{float32} and \code{float64}. Comparison of \code{wceGPU()} from \pkg{survivalGPU} with the reference \code{WCE()} function from the \pkg{WCE} package across \(N = 100\) simulated WCE data sets of \(1{,}000\) patients. Results are reported separately for unconstrained and right-constrained WCE specifications and for models with 1, 2, and 3 interior knots. RE denotes the relative error for the standardized HR contrast defined in Equation~\eqref{eq:wce_re_hr}.}
\label{tab:wce_validation_precision}
\end{table}

The WCE validation results also show strong numerical agreement with the reference implementation. In \code{float32}, the median relative errors for the standardized HR contrast were all below \(8 \times 10^{-5}\), and the largest maximum relative error was \(2.67 \times 10^{-3}\). Thus, even in single precision, the discrepancy with \code{WCE()} remained below \(0.3\%\) in all tested spline specifications. In \code{float64}, the median relative errors were below \(6 \times 10^{-8}\), and all maximum relative errors were below \(3 \times 10^{-6}\). These results show that the \pkg{survivalGPU} WCE implementation reproduces the reference \pkg{WCE} estimates to very high numerical precision in double precision, while the recommended \code{float32} GPU configuration retains small relative errors.

Overall, the validation results support the numerical reliability of \pkg{survivalGPU} for both Cox and WCE models. Double precision provides near-identical agreement with the reference implementations and is therefore useful for validation and sensitivity analyses. Single precision introduces slightly larger numerical differences, as expected, but these differences remained below \(1\%\) for all Cox and WCE validation scenarios. This supports the use of \code{float32} as the recommended high-performance setting for the computational benchmarks and for large-scale GPU analyses.


\section{Benchmarks} \label{sec:benchmark}

The objective of the benchmark study was to assess the computational gain provided by \pkg{survivalGPU} for fitting Cox and WCE models on data sets of increasing size.

\paragraph{Data generation.}
All benchmark data sets were generated using the simulation framework described in Appendix~\ref{sec:simulation}. For all non-null simulation scenarios, the regression coefficient was fixed at $\beta = \log(1.5)$. This value was used only for benchmark data generation and differs from the validation setting in Section~\ref{sec:validation}.

\paragraph{Computational environment.}
All experiments were run on the same workstation, equipped with an Intel Xeon W-2255 CPU running at 3.70 GHz and an NVIDIA GeForce RTX 3090 GPU with 24 GB of memory. CPU computations were restricted to a single core and were not parallelized. The software environment consisted of Ubuntu 24.04.2 LTS, \proglang{Python} 3.13.0, \proglang{R} 4.3.3, \proglang{CUDA} 12.2, \pkg{torch} 2.8.0, \pkg{KeOps} 2.3, \pkg{survival} 3.5.8, \pkg{WCE} 1.0.3, \pkg{lifelines} 0.30.3, \pkg{xgboost} 3.2.0, and \pkg{torchsurv} 0.1.6.

A preliminary warm-up run was performed before timing GPU computations. For each benchmark configuration, the same simulated data set was fitted three times, and the reported runtime corresponds to the mean elapsed wall-clock time across these three runs. Runtime measurements excluded data simulation and package loading.

\paragraph{Floating-point precision.}
All benchmark experiments with \pkg{survivalGPU} were performed using \code{float32} arithmetic. As discussed in Section~\ref{sec:gpu}, single precision is the recommended setting for GPU acceleration because it reduces memory requirements and generally improves runtime, especially on hardware with limited double-precision throughput. The validation study in Section~\ref{sec:validation} reports results for both \code{float32} and \code{float64}, showing that \code{float32} preserves close numerical agreement with the reference implementations while corresponding to the high-performance configuration used in the benchmarks.

\paragraph{Bootstrap runtime estimation.}
For bootstrap experiments, \code{coxphGPU()} and \code{wceGPU()} executed on the GPU were evaluated using their implemented bootstrap procedures. For the CPU backend of \pkg{survivalGPU} and for reference implementations that did not provide parallelized bootstrap support, total runtimes were extrapolated from the runtime of a single model fit. Because a bootstrap analysis with $B$ replicates requires fitting the model once on the original data set and once for each bootstrap sample, the extrapolated runtime was computed as
\begin{equation}
T_{\mathrm{boot}} = (1 + B) T_{\mathrm{fit}},
\label{eq:bootstrap_extrapolation}
\end{equation}
where $T_{\mathrm{fit}}$ is the mean runtime of one model fit over the three repeated runs and $B$ is the number of bootstrap replicates. In all bootstrap benchmarks, the number of bootstrap replicates was set to $B = 1{,}000$.

\subsection[Evaluation of coxphGPU]{\code{coxphGPU()} function} \label{subsec:cox_perf}

\paragraph{Cox model benchmark.}
The GPU-accelerated \code{coxphGPU()} function from \pkg{survivalGPU} was compared with \code{coxph()} from the \pkg{survival} package and with the \code{CoxPHFitter} estimator from the \pkg{lifelines} package in \proglang{Python}. The benchmark used one time-dependent covariate and data sets containing 500 to 1{,}000{,}000 patients. For \code{coxphGPU()}, runtimes are reported for both CPU and GPU execution.

\begin{figure}[H]
\centering
\includegraphics[width=\textwidth]{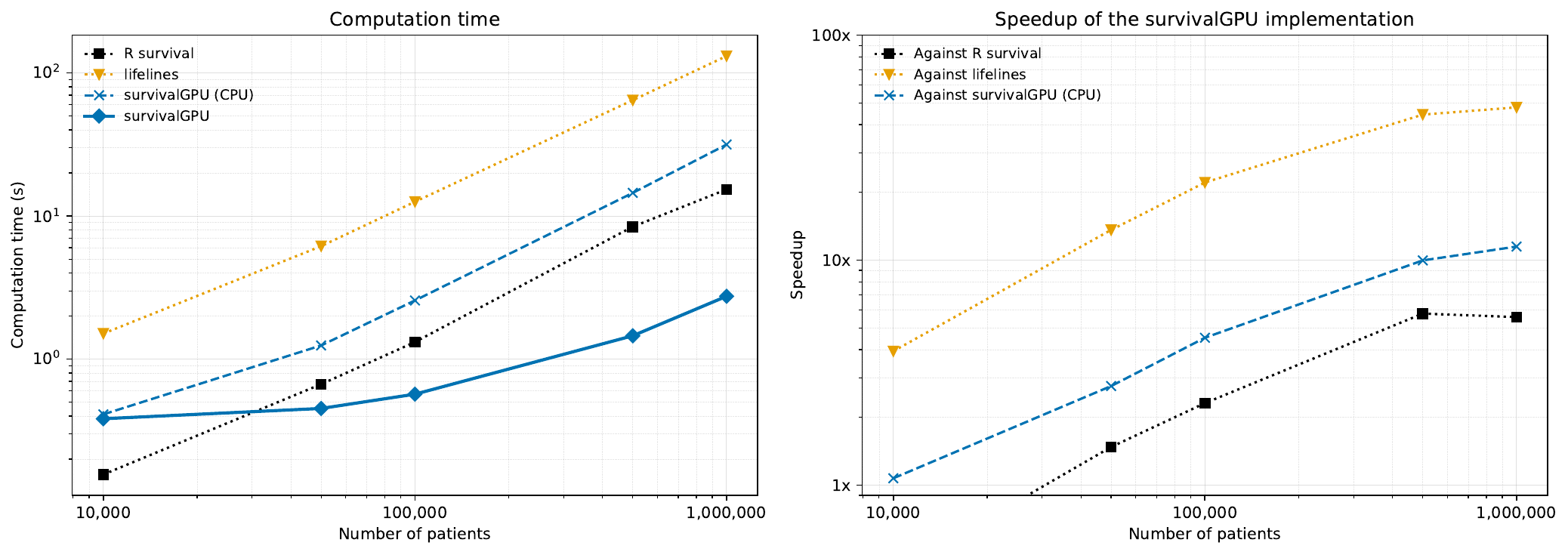}
\caption{Benchmark results for Cox model fitting. Reported runtimes correspond to the mean elapsed time over three runs on the same simulated data set. Left: elapsed time across implementations. Right: speedup of \code{coxphGPU()} with GPU acceleration relative to the other implementations.}
\label{fig:benchmarks_cox}
\end{figure}

Figure~\ref{fig:benchmarks_cox} shows that GPU acceleration reduced the runtime of Cox model fitting for large data sets, although \code{coxph()} from \pkg{survival} was already very efficient. For small data sets, the fixed overhead of GPU execution dominated the computation time. As the sample size increased, however, the GPU version of \code{coxphGPU()} became increasingly advantageous. At $n = 1{,}000{,}000$, \code{coxphGPU()} on the GPU required 2.75 seconds, compared with 15.35 seconds for \code{coxph()} from \pkg{survival}, 31.58 seconds for \code{coxphGPU()} on the CPU, and 131.26 seconds (2.19 minutes) for \code{CoxPHFitter} from \pkg{lifelines}. This corresponded to 5.6-fold, 11.5-fold, and 47.7-fold speedups relative to \pkg{survival}, the CPU backend of \pkg{survivalGPU}, and \pkg{lifelines}, respectively. These results indicate that GPU acceleration provides a clear gain for large data sets, while also showing that standard Cox model fitting is already computationally efficient.

\paragraph{Cox model benchmark with bootstrap resampling.}
We then evaluated runtimes when estimating confidence intervals using $B = 1{,}000$ bootstrap replicates. The Cox bootstrap benchmark was performed up to $n = 100{,}000$ patients. For the CPU backend of \pkg{survivalGPU} and for implementations without parallelized bootstrap support, runtimes were extrapolated according to Equation~\eqref{eq:bootstrap_extrapolation}.

\begin{figure}[H]
\centering
\includegraphics[width=\textwidth]{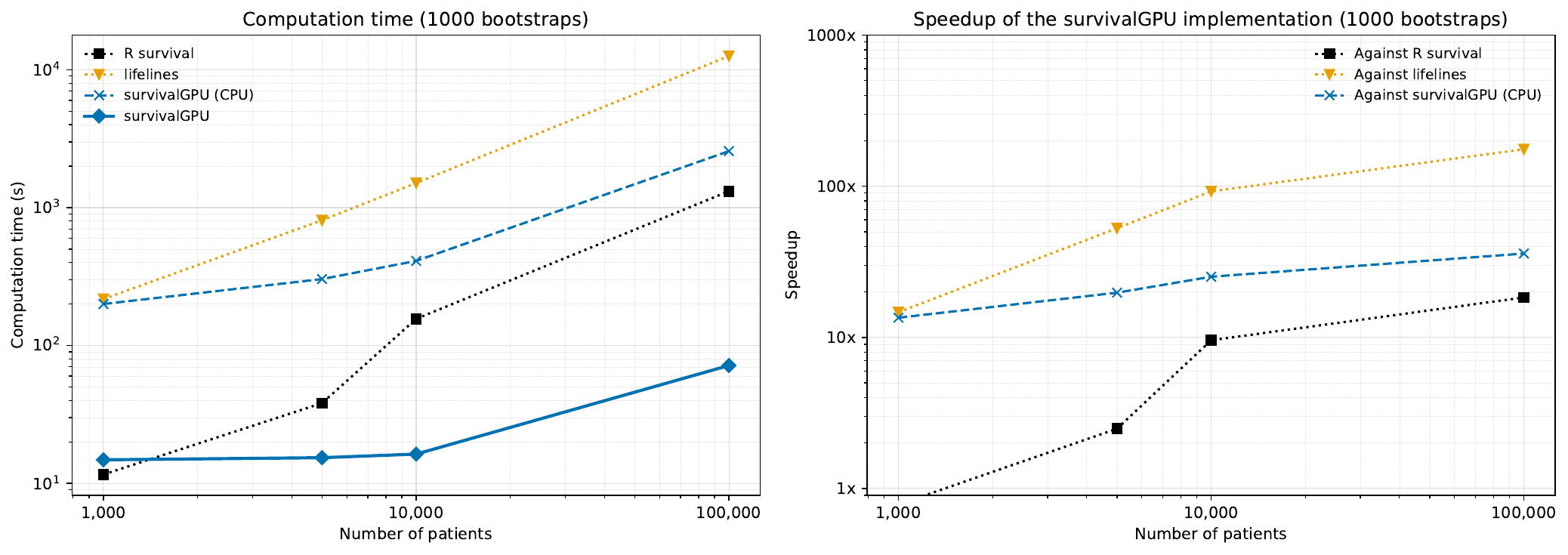}
\caption{Benchmark results for Cox model fitting with $B = 1{,}000$ bootstrap replicates. Reported runtimes correspond to the mean elapsed time over three runs on the same simulated data set. Runtimes for \code{coxphGPU()} on the GPU are measured using the implemented bootstrap procedure, whereas runtimes for the CPU backend of \pkg{survivalGPU} and for implementations without parallelized bootstrap support are extrapolated using Equation~\eqref{eq:bootstrap_extrapolation}. Left: elapsed time across implementations. Right: corresponding speedup; values involving extrapolated runtimes represent projected speedups.}
\label{fig:benchmarks_cox_bootstraps}
\end{figure}

Figure~\ref{fig:benchmarks_cox_bootstraps} shows that bootstrap resampling amplified the advantage of GPU acceleration. For the smallest data sets, the overhead of launching and coordinating the GPU computation remained visible. For larger data sets, however, the parallelized bootstrap implementation in \pkg{survivalGPU} provided a substantial reduction in runtime. At $n = 100{,}000$, \code{coxphGPU()} on the GPU required 71.64 seconds (1.19 minutes), compared with 1{,}004.62 seconds (16.74 minutes) for the extrapolated \pkg{survival} runtime, 2{,}428.25 seconds (40.47 minutes) for the extrapolated runtime of \code{coxphGPU()} on the CPU, and 12{,}697.37 seconds (3.53 hours) for the extrapolated \pkg{lifelines} runtime. This corresponded to 14.0-fold, 33.9-fold, and 177.2-fold speedups relative to \pkg{survival}, the CPU backend of \pkg{survivalGPU}, and \pkg{lifelines}, respectively.

These results highlight the practical benefit of parallelized bootstrap computation for large-scale survival analysis. In particular, GPU acceleration makes bootstrap resampling procedures substantially more practical when confidence intervals must be estimated from a large number of replicates.

\paragraph{Comparison with other GPU-compatible Cox implementations.}
Finally, we compared \code{coxphGPU()} with alternative GPU-compatible approaches related to Cox-type survival objectives. The \pkg{torchsurv} package provides a Cox loss function designed for deep learning frameworks rather than a direct implementation of the Cox proportional hazards model. To obtain a comparable baseline, we combined this loss with an L-BFGS optimizer applied to a linear model with a single layer. The number of optimization iterations was limited to 20, which is consistent with the typical order of magnitude used in standard Cox model fitting.

The \pkg{xgboost} package also provides a survival objective, but does not implement the same Cox model formulation as \code{coxph()} from \pkg{survival} or \code{coxphGPU()} from \pkg{survivalGPU}. For comparability, the number of boosting iterations was restricted to 20, yielding the same nominal number of optimization or boosting steps. This comparison should therefore be interpreted as a computational baseline rather than a strict comparison of statistically identical model-fitting procedures.

\begin{figure}[H]
\centering
\includegraphics[width=0.5\textwidth]{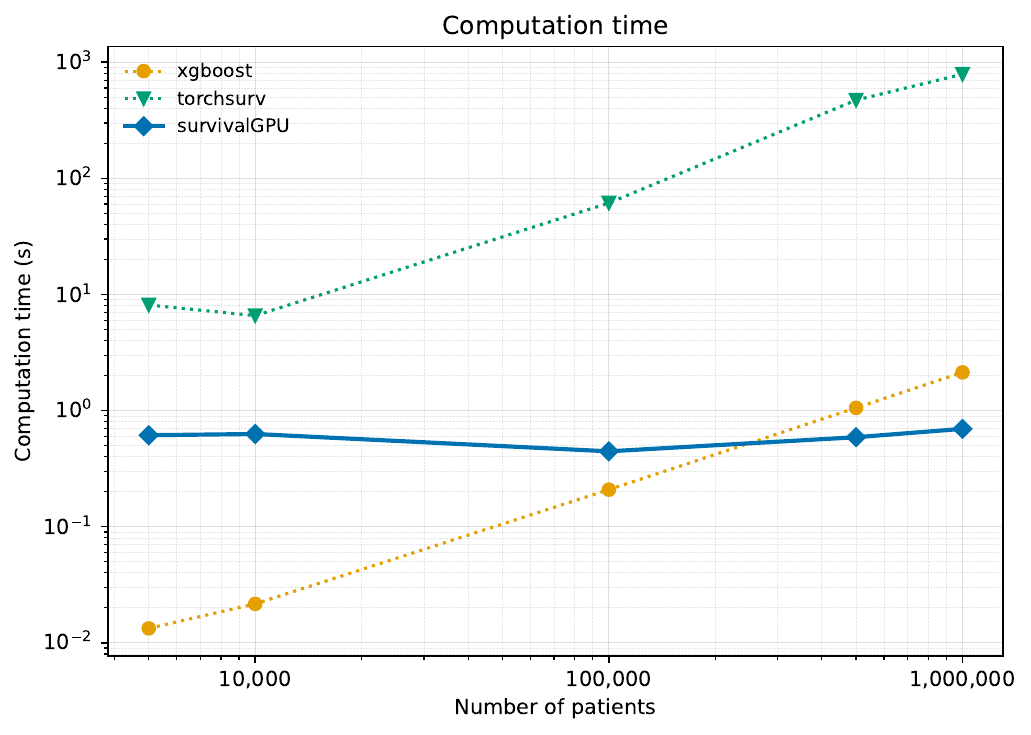}
\caption{Comparison of \code{coxphGPU()} from \pkg{survivalGPU} with alternative GPU-compatible Cox-type survival objectives. Reported runtimes correspond to the mean elapsed time over three runs on the same simulated data set.}
\label{fig:gpu_benchmark}
\end{figure}

Figure~\ref{fig:gpu_benchmark} shows that, for this time-invariant Cox benchmark, \code{coxph()} from \pkg{survival} and the survival objective from \pkg{xgboost} were faster than \code{coxphGPU()} on small data sets, where the fixed overhead of GPU execution dominated the computation time. As the sample size increased, this overhead became negligible, and at the largest sample size \code{coxphGPU()} was the fastest of the tested implementations. At $n = 1{,}000{,}000$, \code{coxphGPU()} required 0.70 seconds, compared with 0.86 seconds for \code{coxph()} from \pkg{survival}, 2.13 seconds for the survival objective from \pkg{xgboost}, and 788.97 seconds (13.15 minutes) for the Cox loss from \pkg{torchsurv}. This corresponded to 1.2-fold, 3.1-fold, and 1{,}134.5-fold speedups relative to \pkg{survival}, \pkg{xgboost}, and \pkg{torchsurv}, respectively. The large difference with \pkg{torchsurv} is expected because \pkg{torchsurv} is primarily designed for integration into deep learning pipelines, where the Cox loss is typically only one component of a larger neural network. Representing time-dependent covariates in \pkg{torchsurv}, when used, is memory-intensive and may lead to GPU memory overflow. The survival objective in \pkg{xgboost} does not support time-dependent covariates. Therefore, this comparison was restricted to time-invariant covariates.

\newpage

\subsection[Evaluation of wceGPU]{Evaluation of the \code{wceGPU()} function} \label{subsec:wce_perf}

\paragraph{WCE model benchmark without bootstrap resampling.}
We then evaluated the runtime required to fit WCE models. The WCE benchmark used data sets ranging from 500 to 100{,}000 patients. For each data set, the WCE model was fitted with a cutoff of 180 days, \code{nknots} set to 3, and an unconstrained WCE specification.

The \code{wceGPU()} function accelerates the two main computational steps of the WCE model. The first step is the construction of the WCE-derived covariates based on the WCE spline basis. This step is the main computational bottleneck in the reference \code{WCE()} implementation from the \pkg{WCE} package. In \pkg{survivalGPU}, this operation is performed using \pkg{KeOps}, which allows efficient symbolic matrix operations without explicitly materializing the full matrices involved in the convolution. The second step is the maximization of the Cox partial likelihood once the WCE-derived covariates have been constructed.

As in the Cox benchmark, we report runtimes for \pkg{survivalGPU} for both CPU and GPU execution. This distinction makes it possible to separate the gain obtained from the \pkg{survivalGPU} implementation on the CPU from the additional gain obtained by executing the computations on the GPU.

\begin{figure}[H]
\centering
\includegraphics[width=\textwidth]{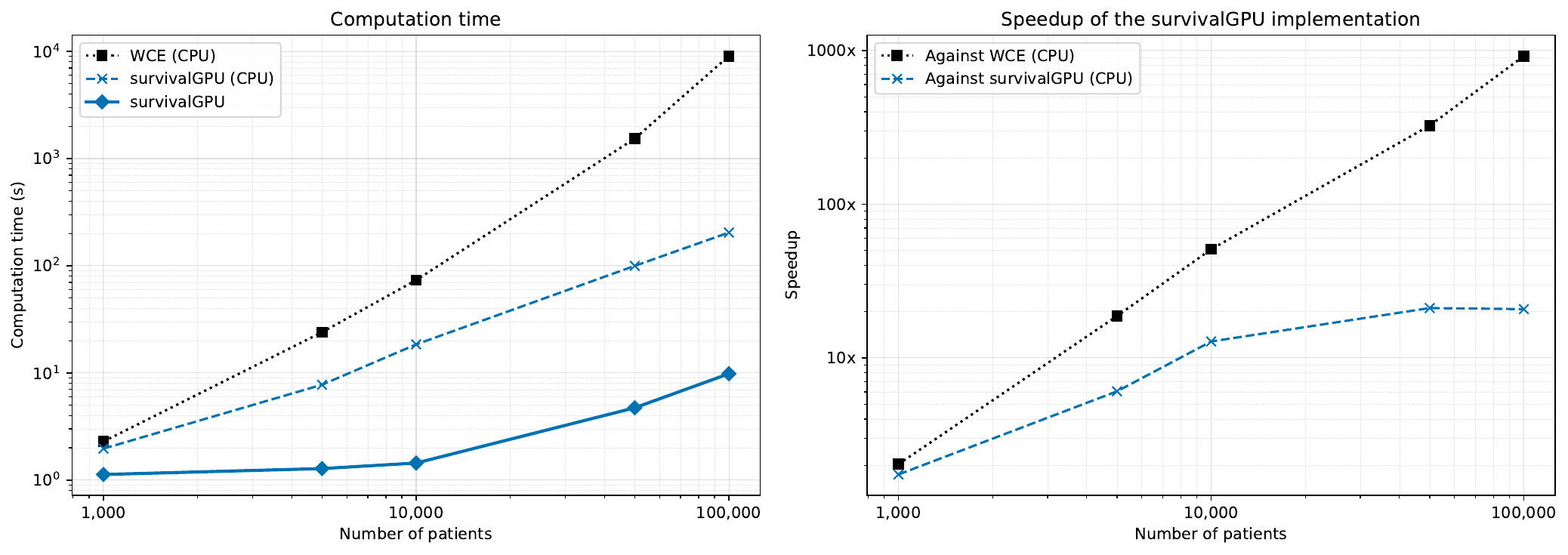}
\caption{Benchmark results for WCE model fitting. Reported runtimes correspond to the mean elapsed time over three runs on the same simulated data set. Left: elapsed time across implementations. Right: speedup of \code{wceGPU()} with GPU acceleration relative to the other implementations.}
\label{fig:benchmarks_wce}
\end{figure}

Figure~\ref{fig:benchmarks_wce} shows that the \pkg{survivalGPU} implementation substantially reduced the runtime of WCE model fitting for moderate and large data sets, both when executed on the CPU and when executed on the GPU. The CPU version of \code{wceGPU()} was already markedly faster than the reference \code{WCE()} implementation for large data sets, reflecting the efficiency of the \pkg{KeOps}-based construction of the WCE-derived covariates. GPU execution provided an additional gain, especially for the largest data sets. At $n = 100{,}000$, \code{wceGPU()} required 9.83 seconds on the GPU, compared with 204.07 seconds (3.40 minutes) on the CPU and 9{,}050.72 seconds (2.51 hours) for the reference \code{WCE()} implementation. This corresponded to 20.8-fold and 920.7-fold speedups relative to the CPU backend of \pkg{survivalGPU} and to \code{WCE()}, respectively.

These results indicate that both components of the implementation contribute to the overall acceleration: the memory-efficient construction of WCE-derived covariates removes the main bottleneck of the reference implementation, while GPU execution further accelerates large-scale model fitting. These gains are important in practice because WCE models are often fitted over a grid of hyperparameter values, including several spline complexities and constraint settings. Reducing the runtime of a single WCE fit therefore directly reduces the cost of model exploration and selection.

\paragraph{WCE model benchmark with bootstrap resampling.}
Finally, we evaluated the runtime of WCE estimation with $B = 1{,}000$ patient-level bootstrap replicates. Bootstrap resampling is commonly used for WCE models to estimate confidence intervals for hazard-ratio contrasts. However, because each bootstrap replicate requires refitting the model, the computational cost can be high for large data sets. Using $B = 1{,}000$ replicates provides a realistic assessment of the computational cost of bootstrap-based uncertainty estimation in WCE analyses.

For the CPU backend of \pkg{survivalGPU} and for implementations that did not provide parallelized bootstrap support, total runtimes were extrapolated using Equation~\eqref{eq:bootstrap_extrapolation} with $B = 1{,}000$. In \pkg{survivalGPU}, the WCE bootstrap procedure avoids recomputing WCE-derived covariates for each bootstrap sample. The WCE-derived covariates are computed once for the original design matrix and reused across bootstrap replicates, after which the Cox partial likelihood is refitted using bootstrap-specific resampling weights.

\begin{figure}[H]
\centering
\includegraphics[width=\textwidth]{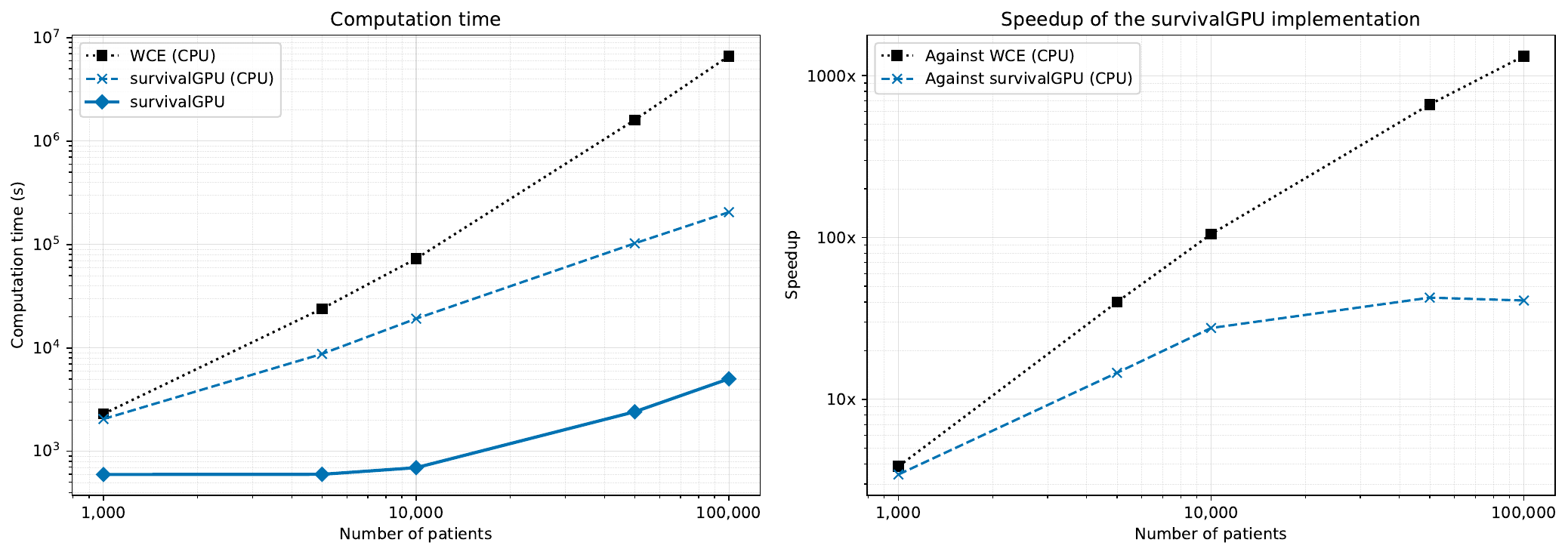}
\caption{Benchmark results for WCE model fitting with $B = 1{,}000$ bootstrap replicates. Reported runtimes correspond to the mean elapsed time over three runs on the same simulated data set. Runtimes for \code{wceGPU()} on the GPU are measured using the implemented bootstrap procedure, whereas runtimes for the CPU backend of \pkg{survivalGPU} and for implementations without parallelized bootstrap support are extrapolated using Equation~\eqref{eq:bootstrap_extrapolation}. Left: elapsed time across implementations. Right: corresponding speedup; values involving extrapolated runtimes represent projected speedups.}
\label{fig:benchmarks_wce_bootstraps}
\end{figure}

Figure~\ref{fig:benchmarks_wce_bootstraps} shows that bootstrap resampling further increased the practical advantage of \pkg{survivalGPU}. The benefit was particularly strong for WCE models because the reference implementation requires repeated construction of WCE-derived covariates and repeated model fitting across bootstrap samples. In contrast, \pkg{survivalGPU} computes the WCE-derived covariates once and reuses them during the bootstrap procedure, substantially reducing the cost of resampling.

At $n = 100{,}000$, \code{wceGPU()} with GPU acceleration completed the bootstrap analysis in 5{,}021.78 seconds (1.39 hours). The extrapolated runtime was 205{,}089.98 seconds (2.37 days) for the CPU backend of \pkg{survivalGPU} and 6{,}635{,}933.69 seconds (76.80 days) for the reference \code{WCE()} implementation. Thus, at this sample size, the GPU implementation was 40.8-fold faster than the extrapolated CPU backend of \pkg{survivalGPU} and 1{,}321.4-fold faster than the extrapolated reference implementation.

The CPU version of \code{wceGPU()} was already substantially faster than the reference implementation, reflecting the efficiency of the \pkg{KeOps}-based construction of the WCE-derived covariates, even when executed on the CPU. However, GPU acceleration provided an additional major gain for bootstrap resampling. Because the benchmark experiments were performed using the recommended \code{float32} GPU configuration described in Section~\ref{sec:gpu}, these results represent the intended high-performance use case of \pkg{survivalGPU}. Together with the numerical validation reported in Section~\ref{sec:validation}, they show that the combination of memory-efficient WCE-derived covariate construction, single-precision GPU arithmetic, and GPU-parallelized model fitting makes bootstrap estimation of confidence intervals feasible for large WCE analyses that would otherwise be computationally prohibitive.


\section{Illustration}
\label{sec:empirical}

To illustrate the use of the \pkg{survivalGPU} package, we analyze the \code{drugdata} dataset from the \pkg{WCE} package. The data contain follow-up information for 500 patients, including time-varying drug exposure through the \code{dose} variable, event indicators, and baseline covariates such as \code{age} and \code{sex}. Note that the dataset contains one line per unit of follow-up time, defined by the \code{Start} and \code{Stop} variables, rather than one line per interval over which the covariates remain constant. 

The objective of this example is not to draw substantive conclusions about a real drug effect, but to demonstrate a typical workflow using \pkg{survivalGPU} for the \code{coxphGPU()} and \code{wceGPU()} functions in \proglang{R}.

\subsection[Analysis of the drugdata dataset with the coxphGPU() function]{Analysis of the \code{drugdata} dataset with the \code{coxphGPU()} function} \label{subsec:cox_illustration}

A Cox model considers each line of the dataset to be an independent patient in calculations. As the \code{drugdata} dataset contains one line per unit of time, it is not optimal for the Cox model. It is better to use a function like \code{survcondense} of the \pkg{survival} package to compress the dataset so it has only one line per interval over which the covariates remain constant. It will not change the results of the analysis but the computations will be quicker.

\begin{CodeChunk}
\begin{CodeInput}
R> data("drugdata",package="WCE")
R> compressed_drugdata <- survcondense(
R>   Surv(Start, Stop, Event) ~ age + sex + dose,
R>   data = drugdata,
R>   id   = Id
R> )
R>
R> print(head(compressed_drugdata))

\end{CodeInput}
\begin{CodeOutput}
  Id age sex dose Start Stop Event
1  1  40   1  0.0     0   14     0
2  1  40   1  2.0    14   21     0
3  1  40   1  0.0    21   28     0
4  1  40   1  2.5    28   49     1
5  2  48   0  0.0     0   14     0
6  2  48   0  1.5    14   63     0
\end{CodeOutput}
\end{CodeChunk}

\code{coxphGPU} is used almost like \code{coxph} from the \pkg{survival} package. The bootstraps are done using the \code{bootstrap()} function on the fitted Cox object. To use the bootstrap function it's necessary to save the design matrix of the data used by the model with the \code{x = TRUE} argument. It's also necessary to add the \code{patient_id} of the model, here the column \code{"Id"}. Without it, the bootstrap function wouldn't be able to resample on different patients and the results would be meaningless. The \code{batchsize} argument sets the number of bootstraps run in parallel on the GPU. The \code{summary} is identical to the one in the \pkg{survival} package; however, if bootstraps were run, it adds the confidence intervals calculated with the bootstraps.

\begin{CodeChunk}
\begin{CodeInput}
R> result <- coxphGPU(Surv(Start, Stop, Event) ~ dose + age + sex,
+                 data = compressed_drugdata, 
+                 x = TRUE, patient_id = "Id", device ="gpu") 
R> result_boot <- bootstrap(result, R = 100, batchsize = 25)
R> summary(result)
\end{CodeInput}
\begin{CodeOutput}
         coef exp(coef) se(coef)     z Pr(>|z|)    
dose 0.226072  1.253666 0.047500 4.759 1.94e-06 ***
age  0.011103  1.011165 0.003966 2.800  0.00511 ** 
sex  0.627285  1.872519 0.117760 5.327 1.00e-07 ***
---
Signif. codes:  0 ‘***’ 0.001 ‘**’ 0.01 ‘*’ 0.05 ‘.’ 0.1 ‘ ’ 1

     exp(coef) exp(-coef) lower .95 upper .95
dose     1.254     0.7977     1.142     1.376
age      1.011     0.9890     1.003     1.019
sex      1.873     0.5340     1.487     2.359

Concordance= 0.619  (se = 0.017 )
Likelihood ratio test= 55.44  on 3 df,   p=6e-12
Wald test            = 58.46  on 3 df,   p=1e-12
Score (logrank) test = 59.94  on 3 df,   p=6e-13

 ---------------- 
Confidence interval with 100 bootstraps for exp(coef), conf.level = 0.95 :
        2.5%   97.5%
dose 1.12260 1.36806
age  1.00466 1.02059
sex  1.50640 2.36081

\end{CodeOutput}
\end{CodeChunk}

The dose, the age and the sex all have a p-value of less than 0.05, so they are all significant. We can see that the confidence intervals estimated by the model and by the bootstraps are very similar. 

The function outputs a \code{coxphGPU} object constructed by S3 inheritance from the \code{coxph} object of the \pkg{survival} package. That means that all the functions of the package that can be used on a \code{coxph} object can be used on our result. Here, for example, we can test the proportional hazards assumption. 

\begin{CodeChunk}
\begin{CodeInput}
R> cox.zph(result)
\end{CodeInput}
\begin{CodeOutput}
         chisq df    p
dose   2.08931  1 0.15
age    0.37223  1 0.54
sex    0.00116  1 0.97
GLOBAL 2.53698  3 0.47
\end{CodeOutput}
\end{CodeChunk}

No p-value is lower than 0.05, so the proportional hazards assumption is respected for all the covariates. 

And as our \code{coxphGPU} object inherits the properties of the \code{coxph} object, we can also use the functions of other packages that work on a \code{coxph} object. For example, we can plot the adjusted survival curves from the Cox model using the \pkg{survminer} package.

\begin{CodeChunk}
\begin{CodeInput}
R> library("survminer")
R> ggadjustedcurves(result, variable = "sex", data = drugdata)
\end{CodeInput}
\begin{CodeOutput}


\end{CodeOutput}
\end{CodeChunk}


\subsection[Analysis of the drugdata dataset with the wceGPU() function]{Analysis of the \code{drugdata} dataset with the \code{wceGPU()} function } \label{subsec:cox_illustration}

In contrast to the Cox model, the WCE model cannot be fitted on a data set compressed into intervals over which the covariates remain constant. The hazard at time $u$ depends on the full exposure history through the WCE-derived covariates defined in Equation~\ref{eq:gpu_wce_feature}. These covariates change at every follow-up time, even during periods when the dose itself is constant. The data set must therefore contain one row per unit of follow-up time for each patient. The \code{drugdata} data set is already in this format, so no transformation is needed. For data sets recorded as intervals of constant exposure, the \code{survSplit()} function from the \pkg{survival} package can be used to expand them into one row per unit of follow-up time.

\begin{CodeChunk}
\begin{CodeInput}
R> data("drugdata", package = "WCE")
R> print(head(drugdata))
\end{CodeInput}
\begin{CodeOutput}
  Id Event Start Stop sex age dose
1  1     0     0    1   1  40    0
2  1     0     1    2   1  40    0
3  1     0     2    3   1  40    0
4  1     0     3    4   1  40    0
5  1     0     4    5   1  40    0
6  1     0     5    6   1  40    0
\end{CodeOutput}
\end{CodeChunk}

Fitting a WCE model requires choosing three hyperparameters, described in Section~\ref{subsec:wce_model}: the cutoff, the number of interior knots and the boundary constraint.

The cutoff defines the time window over which past exposures are assumed to affect the current hazard, and should be chosen from prior knowledge of the drug. If the cutoff is too short, the estimated weight function cannot capture the full effect of past exposures. If it is too long, the spline knots are spread over a wider interval, and short-term effects are estimated less accurately. In this example, the cutoff is set to 180 days.

The \code{nknots} argument sets the number of interior knots of the cubic B-spline basis. Additional knots make the estimated weight function more flexible but increase the risk of overfitting. When a vector of candidate values is given to \code{nknots}, \code{wceGPU()} fits one model for each value and retains the model with the lowest BIC, defined in Equation~\eqref{eq:BIC}.

The \code{constrained} argument determines whether the estimated weight function is forced to reach zero smoothly at the cutoff. If the cutoff is long enough to contain the full effect of past exposures, the true weight function is close to zero at the cutoff and the right constraint is appropriate. If the cutoff is too short, the true weight function is still nonzero at the cutoff, but the constraint forces the estimate to zero regardless. To assess whether the chosen cutoff is long enough, the model is fitted with and without the right constraint. If the unconstrained model has the lower BIC, it fits the data better, which suggests that the cutoff is too short. The \code{summary()} function give the result of the fitted model. 

\begin{CodeChunk}
\begin{CodeInput}
R> fit <- wceGPU(
+  data = drugdata, id = "Id", event = "Event", start = "Start",
+  stop = "Stop", expos = "dose", covariates = c("age", "sex"),
+  nknots = 1:3, cutoff = 180, constrained = "Right", x = TRUE
+)
R> summary(fit)
\end{CodeInput}
\begin{CodeOutput}
Weight function: cubic spline with 1 interior knot(s), constrained to go smoothly to zero for exposures far in the past

Covariates:
         coef exp(coef)  se(coef)      z  Pr(>|z|)    
age 0.0118744 1.0119451 0.0039891 2.9767  0.002914 **
sex 0.7054766 2.0248115 0.1191675 5.9200 3.219e-09 ***
---
Signif. codes:  0 ‘***’ 0.001 ‘**’ 0.01 ‘*’ 0.05 ‘.’ 0.1 ‘ ’ 1

Partial log-likelihood: -1887.195, BIC: 3804.131
Number of events: 383

See plot(fit) for the estimated weight function.
\end{CodeOutput}
\end{CodeChunk}

Here the model corresponding to the lowest BIC is the model with one interior knot. It has a BIC of around 3804. It the model with a Right constraint. When launching the model without constraint, the BIC is 3812. It's bigger than 3804 so that means that the cutoff is not too short. 

Here the age and sex covaraites are both significative. However we don't have any information on the effect of the dose covariate modelled with the exposure. First we can see the coef for the wce-derived covaraites with the \code{coef()} function.

\begin{CodeChunk}
\begin{CodeInput}
R> coef(result)
\end{CodeInput}
\begin{CodeOutput}
\$est
             D1          D2          D3         D4         D5
[1,] 0.03202082 0.002040809 0.005966648 0.01768173 -0.0216031

\$covariates
            age       sex
[1,] 0.01186158 0.7045086
\end{CodeOutput}
\end{CodeChunk}

The \code{coef()} method returns the estimated coefficients of the WCE-derived covariates and of the adjustment covariates. The largest coefficient corresponds to the first WCE-derived covariate, which suggests a strong short-term effect. However, the spline coefficients are difficult to interpret individually, because the shape of the weight function results from the combination of all basis functions. The estimated weight function can instead be visualized using the \code{plot()} method.

\begin{CodeChunk}
\begin{CodeInput}
R> plot(model)
\end{CodeInput}
\end{CodeChunk}

\begin{figure}[H]
\centering
\includegraphics[width=0.55\textwidth]{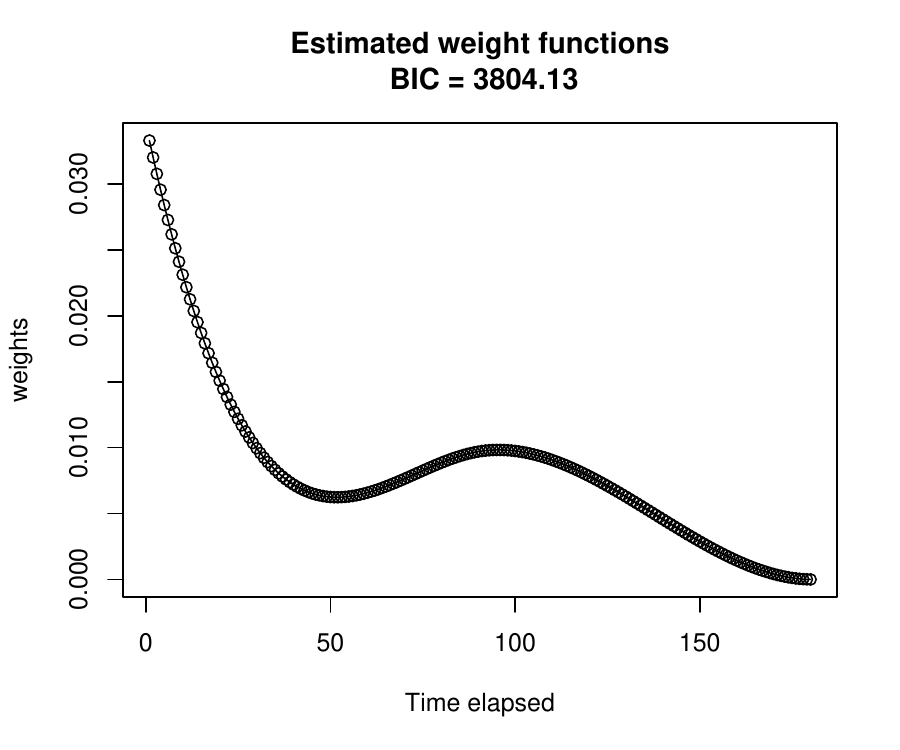}
\caption{Estimated WCE weight function for the selected model fitted to the \code{drugdata} data set. The exposure cutoff is fixed at 180 days, and the model uses one interior knot and a right-constrained weight function.}
\label{fig:wceplot}
\end{figure}

Figure~\ref{fig:wceplot} shows that the most recent exposures receive the largest weights. The weights decrease sharply over the first 50 days, remain at a lower and roughly stable level until about 150 days, and then decrease to zero at the cutoff, as imposed by the right constraint.

Because the WCE metric depends on the full exposure history, the effect of the drug cannot be summarized by a single coefficient. Instead, hazard ratios are obtained by comparing two exposure histories, as defined in Equation~\eqref{eq:hr_wce}. A simple contrast compares a patient exposed to one dose unit throughout the cutoff window with a patient who remains unexposed over the same window. This hazard ratio is computed with the \code{HR()} function.

\begin{CodeChunk}
\begin{CodeInput}
R> exposed   <- rep(1, cutoff)
R> unexposed <- rep(0, cutoff)
R> HR(fit, exposed, unexposed)
\end{CodeInput}
\begin{CodeOutput}
           HR
[1,] 4.534537
\end{CodeOutput}
\end{CodeChunk}

The estimated hazard ratio is 4.53, which indicates a strong association between exposure and the event. However, this estimate comes without a confidence interval. Wald-type inference on the WCE-derived covariates would be difficult to interpret, because it yields one p-value and one confidence interval for each spline coefficient rather than for the hazard ratio. It would also ignore the uncertainty introduced by the selection of the number of knots.

Confidence intervals for the hazard ratio are therefore obtained by patient-level bootstrap resampling, using the \code{bootstrap()} function. All candidate numbers of knots are fitted on each bootstrap sample, and the number of knots is selected again in each replicate, so that the confidence interval accounts for the uncertainty in this choice. As for the Cox model, the \code{R} argument sets the number of bootstrap replicates and the \code{batchsize} argument sets the number of replicates processed in parallel on the GPU.

\begin{CodeChunk}
\begin{CodeInput}
R> fit_boot <- bootstrap(fit, R = 100, batchsize = 25)
R> HR(fit_boot, exposed, unexposed)
\end{CodeInput}
\begin{CodeOutput}
           HR  CI 2.5% CI 97.5%
[1,] 4.534537 3.345773 6.549886
\end{CodeOutput}
\end{CodeChunk}

The 95\% bootstrap percentile confidence interval ranges from 3.35 to 6.55. Because it excludes 1, the association between exposure and the event is statistically significant at the 5\% level. Since \code{drugdata} is a simulated data set, this result illustrates the output of \pkg{survivalGPU} rather than providing evidence about a real drug effect.

\section{Conclusion and future work} \label{sec:conclusion}

\pkg{survivalGPU} provides a GPU-accelerated implementation of Cox and weighted cumulative exposure (WCE) models. It extends existing survival analysis software by combining compatibility with established \proglang{R} workflows and a \proglang{Python} interface. The package reproduces the results of the Cox routines from \pkg{survival} and of the \code{WCE()} function from the \pkg{WCE} package within precision-specific numerical tolerances in both \code{float32} and \code{float64}, while substantially reducing computation time for large data sets and bootstrap-based estimation of confidence intervals.

The largest gains were observed when computations involved repeated model fitting. For the Cox model, GPU acceleration made \code{coxphGPU()} \(5.6\times\) faster than \code{coxph()} on 1{,}000{,}000 patients. With \(B = 1{,}000\) bootstrap replicates on 100{,}000 patients, \code{coxphGPU()} required 71.64 seconds (1.19 minutes), compared with an extrapolated runtime of 1{,}004.62 seconds (16.74 minutes) for \code{coxph()}, corresponding to a \(14.0\times\) speedup. For WCE models, the gains were larger: with 100{,}000 patients, \code{wceGPU()} required 9.83 seconds for a single model fit, compared with 9{,}050.72 seconds (2.51 hours) for the reference \code{WCE()} implementation, corresponding to a \(920.7\times\) speedup. With \(B = 1{,}000\) bootstrap replicates, \code{wceGPU()} required 5{,}021.78 seconds (1.39 hours), compared with an extrapolated runtime of 6{,}635{,}933.69 seconds (76.80 days) for \code{WCE()}, corresponding to a \(1{,}321.4\times\) speedup.

These speedups broaden the practical use of WCE models. Analyses that would previously require hours or days can be reduced to minutes, or seconds for single model fits, making it feasible to fit models over grids of hyperparameters, compare alternative constraints, and compute bootstrap confidence intervals on large cohorts. This is particularly important in pharmacoepidemiology, where exposure histories are longitudinal, often sparse, and may involve large cohorts.

A central component of these gains is the use of \pkg{KeOps} to compute the WCE-derived covariates used in the WCE model. This step represents the main bottleneck of the reference implementation. In \pkg{survivalGPU}, the memory-efficient formulation already provides substantial acceleration when executed on the CPU, while GPU execution adds a further speedup, especially for large data sets and bootstrap-based estimation of confidence intervals.

The design of \pkg{survivalGPU} emphasizes flexibility, extensibility, and long-term maintainability. The core implementation is written in \proglang{Python} and follows conventions used by \pkg{scikit-learn}, making it suitable for modern \proglang{Python}-based data science pipelines. At the same time, the \proglang{R} interface preserves a syntax familiar from \code{coxph()} in \pkg{survival} and \code{WCE()} in \pkg{WCE}, allowing the package to fit seamlessly into existing statistical and survival analysis workflows.

Future work will extend the current implementation to more complex WCE settings, including multiple concurrent exposures and exposure interactions. Because such extensions may increase model dimension, regularization methods will be incorporated to control overfitting and improve numerical stability. Additional Cox model extensions, including penalization, stratification, competing risks \citep{fine1999proportional}, and automated variable selection \citep{tibshirani1997lasso}, will also be considered.

These developments will be evaluated through simulation studies and applied to real-world pharmacoepidemiological data, in particular large healthcare databases such as the SNDS accessed through the Health Data Hub infrastructure. In this setting, \pkg{survivalGPU} may help make flexible WCE modeling more practical for the study of adverse drug reactions in patients exposed to complex treatment histories.


\section*{Acknowledgments}

This work benefited from state aid managed by the Agence Nationale de la Recherche under France 2030, under reference “ANR-23-IACL-0008” (PR[AI]RIE-PSAI). It was also funded by EPI-PHARE under the 2021 call for projects.


\bibliography{refs}

\appendix

\section{Data simulation} \label{sec:simulation}

\subsection{The permutational algorithm} \label{subsec:permalgo}

The simulated data sets used in this study were generated using the permutational algorithm, originally introduced by \citet{mackenzie_marginal_nodate} for simulating time-to-event data under the Cox model framework and later adapted to the WCE model by \citet{sylvestre_flexible_2009}. In the present work, we implemented a version based on the \code{torch} module in \pkg{PyTorch}, allowing the simulation step to be executed on the GPU, and used it to generate simulated data sets for validation and benchmarking.

The algorithm takes \(n\) covariate trajectories, a set of observed follow-up times and event indicators, and a scenario defined by regression coefficients as inputs. For WCE simulations, it also takes an exposure trajectory \(X_i(t)\), a weight function \(w(\cdot)\), and a WCE coefficient \(\beta_{\mathrm{WCE}}\). For Cox simulations, the WCE term is omitted and the linear predictor depends only on Cox covariates.

The observed follow-up times are sorted in increasing order and matched sequentially to the covariate or exposure trajectories. At censoring times, an individual is sampled uniformly from the current risk set. At event times, an individual is sampled from the current risk set with probability proportional to \(\exp\{\eta_i(t)\}\), where \(\eta_i(t)\) is the linear predictor of the Cox or WCE model. The selected individual is then removed from all subsequent risk sets. In Cox-only scenarios, the WCE term is set to zero; when no additional Cox covariates are included, the Cox covariate term is set to zero. The detailed steps are presented in Algorithm~\ref{alg:permalgo}.

\begin{algorithm}[h!]
\caption{Permutational algorithm for Cox and WCE simulations}
\label{alg:permalgo}
\begin{algorithmic}[1]
\State \textbf{Input:}
  \begin{itemize}
    \item \(n\) observed follow-up times \(t_1,\ldots,t_n\) and event indicators \(\delta_1,\ldots,\delta_n\), where \(\delta_i = 1\) denotes an event and \(\delta_i = 0\) denotes censoring;
    \item model-specific covariate trajectories, including at least one of the following:
    \begin{itemize}
      \item exposure trajectories \(\mathbf{X}_i = \{X_i(0), \ldots, X_i(t_{\max})\}\), a normalized weight function \(w(\cdot)\) defined on \([0,a]\), and a WCE coefficient \(\beta_{\mathrm{WCE}}\);
      \item Cox covariates \(\mathbf{Z}_i(t) = (Z_{i1}(t), \ldots, Z_{iq}(t))\), fixed or time-dependent, and their coefficients \(\boldsymbol{\alpha} = (\alpha_1, \ldots, \alpha_q)\).
    \end{itemize}
  \end{itemize}

\State \textbf{Output:}
  \begin{itemize}
    \item \(n\) individuals with covariate and exposure trajectories matched to observed follow-up times and event indicators.
  \end{itemize}

\State If the scenario includes a WCE exposure, compute the WCE metric
\[
\mathrm{WCE}_i(u) =
\sum_{\tau=0}^{\min(a,u)} w(\tau) X_i(u-\tau),
\quad \forall i,u.
\]

\State Sort observed follow-up times and indicators:
\[
(t_{(1)}, \delta_{(1)}), \ldots, (t_{(n)}, \delta_{(n)})
\quad \text{with} \quad
t_{(1)} \leq \cdots \leq t_{(n)}.
\]

\For{\(k = 1\) to \(n\)}
  \State Identify the risk set \(R(t_{(k)})\), defined as the individuals not yet assigned to a previous observed time.

  \If{\(\delta_{(k)} = 0\)}
    \State Draw individual \(i_k\) uniformly at random from \(R(t_{(k)})\).
    \State Assign censoring time \(t_{(k)}\) to individual \(i_k\).
  \Else
    \State For all \(j \in R(t_{(k)})\), compute
    \[
    P_j(t_{(k)}) =
    \frac{
    \exp\left\{
      \beta_{\mathrm{WCE}} \mathrm{WCE}_j(t_{(k)})
      +
      \boldsymbol{\alpha}^{\top}\mathbf{Z}_j(t_{(k)})
    \right\}
    }{
    \sum_{\ell \in R(t_{(k)})}
    \exp\left\{
      \beta_{\mathrm{WCE}} \mathrm{WCE}_\ell(t_{(k)})
      +
      \boldsymbol{\alpha}^{\top}\mathbf{Z}_\ell(t_{(k)})
    \right\}
    }.
    \]
    \State Draw individual \(i_k\) from \(R(t_{(k)})\) with probabilities \(P_j(t_{(k)})\).
    \State Assign event time \(t_{(k)}\) to individual \(i_k\).
  \EndIf

  \State Remove \(i_k\) from all subsequent risk sets.
\EndFor
\end{algorithmic}
\end{algorithm}

\subsection{Simulation framework} \label{subsec:simulation_framework}

\paragraph{General framework.}
The simulation framework used in this work closely follows the design of \citet{sylvestre_flexible_2009}, which serves as a reference for the validation of the WCE model. The same permutational framework was also used to generate Cox model scenarios with fixed or time-dependent covariates.

Each patient was followed for at most 365 days. Marginal event times were generated independently of exposure and covariates from a uniform distribution on \([0,365]\) days. Random censoring times were generated independently from a uniform distribution on \([0,730]\) days, yielding approximately 50\% censoring. The observed follow-up time was defined as the minimum of the event and censoring times, and the event indicator was equal to one if the event occurred before censoring.

\paragraph{WCE exposure trajectories.}
For WCE simulations, treatment exposure varied both between individuals and over time within individuals. Patients alternated between periods of treatment and periods of interruption. The duration of each treatment or interruption period was generated from a log-normal distribution with meanlog \(= \log(0.5)\) and sdlog \(= 0.8\), expressed in weeks and rounded up to the nearest week. This alternating process was repeated until the end of follow-up.

Within each uninterrupted treatment period, a constant daily dose was randomly assigned from the set \(\{0.5, 1, 1.5, 2, 2.5, 3\}\) and remained fixed throughout the period. During treatment interruptions, the dose was set to zero.

\paragraph{Cox covariate scenarios.}
For Cox simulations with a fixed covariate, each individual was assigned a binary covariate \(Z_i\) with equal probabilities for the two groups. The event probability was then driven by the Cox linear predictor \(\alpha Z_i\), where \(\alpha\) is the log-hazard ratio specified for the simulation scenario.

For Cox simulations with a time-dependent covariate, the covariate trajectory \(Z_i(t)\) was generated using the same intermittent mechanism as for the WCE exposure trajectories. However, in contrast to the WCE scenario, the current value \(Z_i(t)\) entered the Cox linear predictor directly through \(\alpha Z_i(t)\), without convolution with a weight function.

For scenarios with ten covariates, this generation mechanism was repeated independently for each covariate, and the first coefficient was used for numerical validation. The same coefficient value was assigned to each simulated covariate.

The permutational algorithm was then used to match the observed follow-up times and event indicators to the simulated covariate and exposure trajectories according to the model-specific linear predictor.

\begin{figure}[H]
\centering
\includegraphics[width=1\textwidth]{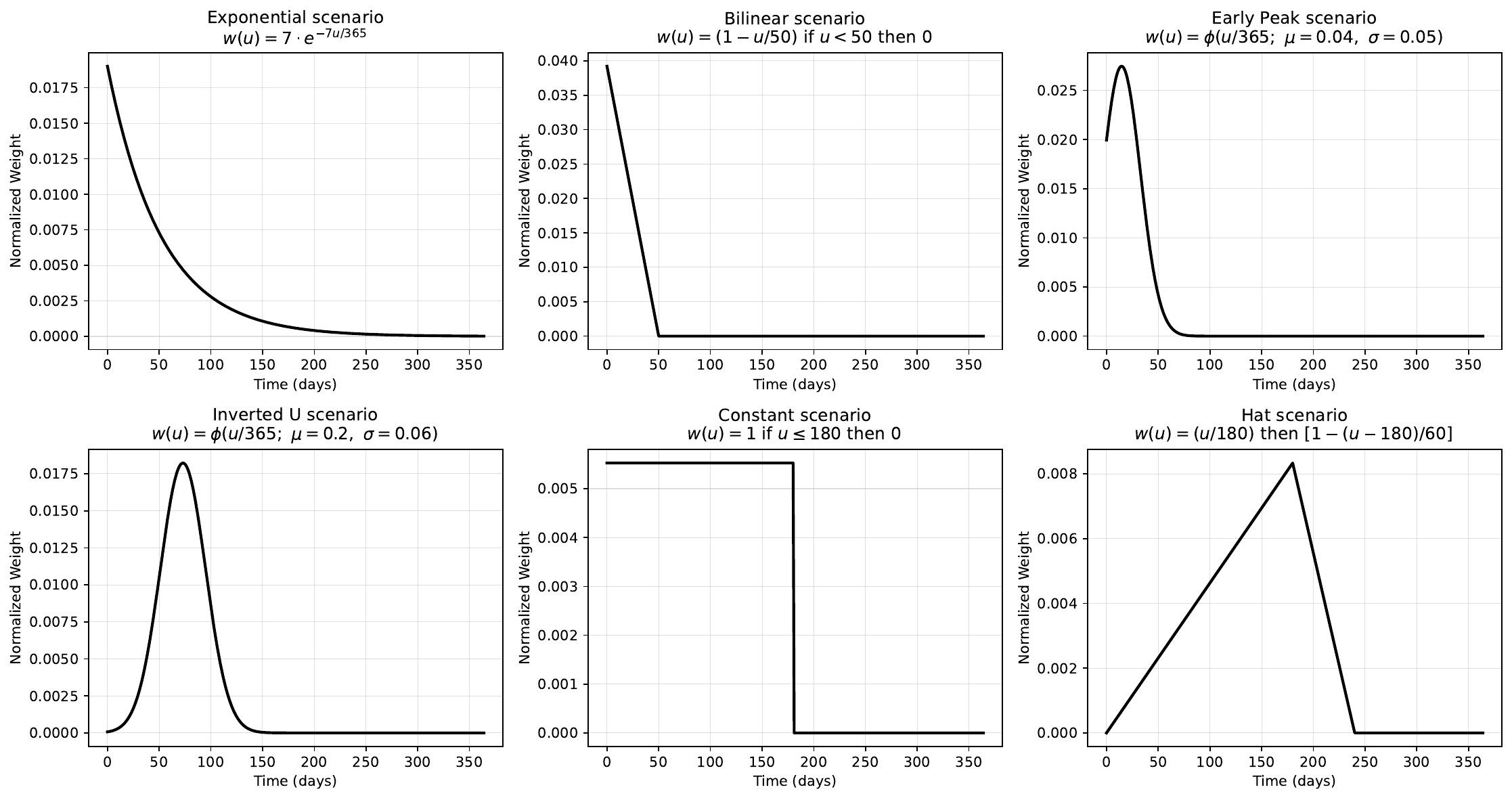}
\caption{Illustration of the six non-null weight functions used in the WCE simulation scenarios. Each function is normalized so that the area under its curve is equal to 1.}
\label{fig:wce_shapes}
\end{figure}

\paragraph{Weight-function scenarios.}
Six non-null weight functions \(w(\tau)\) were used to evaluate numerical agreement across different weight-function shapes. Each non-null function was defined on the lag interval \([0,180]\) days, normalized so that \(\sum_{\tau=0}^{180} w(\tau) = 1\), and set to zero for lags larger than 180 days. These scenarios represent different assumptions about how the effect of exposure changes with time since exposure, including monotonically decreasing, delayed, and approximately constant effects. The shapes of the weight functions are shown in Figure~\ref{fig:wce_shapes}.

An additional null scenario was included by setting \(w(\tau)=0\) for all lags, so that exposure had no effect on the hazard. This null scenario was not normalized, since exposure was defined to have no effect.

\end{document}